\documentclass[11pt,a4paper]{article}
\usepackage{cite}
\usepackage{appendix}
\usepackage{caption}
\usepackage{indentfirst}
\usepackage[utf8]{inputenc}
\usepackage[T1]{fontenc}
\usepackage{cancel,xcolor}
\usepackage{ulem}
\usepackage{dsfont}
\usepackage{amsmath}
\usepackage{amssymb}
\usepackage{amsfonts}
\usepackage{amsthm}
\usepackage{float}
\usepackage[final]{pdfpages}
\usepackage{graphicx}
\usepackage[scale=0.8]{geometry}
\usepackage[english]{babel}
\usepackage{tcolorbox}
\usepackage{url}
\usepackage{hyperref}
\usepackage{empheq}
\usepackage[font=small,labelfont=bf]{caption}
\usepackage{textcomp}
\usepackage{setspace}
\newcommand{\ii}{{\rm i}}
\newcommand{\dd}{{\rm d}}
\newcommand{\ee}{{\rm e}}

\definecolor{carmine}{rgb}{0.59, 0.0, 0.09}
\definecolor{darkblue}{rgb}{0.0, 0.0, 0.55}
\definecolor{burntorange}{rgb}{0.8, 0.33, 0.0}
\definecolor{dark-green}{RGB}{0, 128, 0}

\begin{document}

\title{Topology of shallow-water waves on the rotating sphere}
\author{Nicolas Perez$^1$, Armand Leclerc$^1$, Guillaume Laibe$^1$ and Pierre Delplace$^2$}
\date{}

\maketitle

\begin{center}
$^1$\textit{Univ Lyon, ENS de Lyon, Univ Claude Bernard, CNRS, Centre de Recherche Astrophysique de Lyon (UMR CNRS 5574), F-69230 Saint-Genis-Laval, France}
\end{center}

\begin{center}
$^2$\textit{Univ Lyon, ENS de Lyon, Univ Claude Bernard, CNRS, Laboratoire de Physique (UMR CNRS 5672), F-69342 Lyon, France}
\end{center}

\begin{center}
email: \texttt{nicolas.perez@ens-lyon.fr}
\end{center}

\begin{abstract}
Topological properties of the spectrum of shallow-water waves on a rotating spherical body are established. Particular attention is paid to its spectral flow, i.e. the modes whose frequencies transit between the Rossby and inertia-gravity wavebands as the zonal wave number is varied. Organising the modes according to the number of zeros of their meridional velocity, we conclude that the net number of modes transiting between the shallow-water wavebands on the sphere is null, in contrast with the Matsuno spectrum. This difference can be explained by a miscount of zeros under the $\beta$-plane approximation. We corroborate this result with the analysis of \cite{delplace2017} by showing that the curved metric discloses a pair of degeneracy points in the Weyl symbol of the wave operator, non-existent under the $\beta$-plane approximation, each of them bearing a Chern number $-1$.
\end{abstract}

\section{Introduction} \label{sec:intro}

The rotating shallow-water model is certainly one of the most significant for modeling two-dimensional large-scale fluid motions in the ocean, the atmosphere and even stellar media \cite{gill1982atmosphere,vallis2017atmospheric,zeitlin2018geophysical,gilman2000magnetohydrodynamic,zaqarashvili2009,zaqarashvili2021rossby}. It was originally introduced by Laplace in 1775 to address the problem of the dynamical response of the oceans to the tidal forces generated by the Moon. The strength of this model relies on the fact that it focuses mainly on the effects of the Coriolis force, produced by the solid-body rotation of the planet or star at rate $\mathbf{\Omega}$, on the dynamics of surface waves, without losing so much generality. Owing to the curvature of the surface, these Coriolis effects, i.e. the projection of the Coriolis term $-2 \mathbf{\Omega} \times \mathbf{v}$ on the surface (with $\mathbf{v}$ the fluid velocity field on this surface), naturally depend on the latitude $\theta$, more specifically on the normal component of $2 \mathbf{\Omega}$, called the \textit{Coriolis parameter}, which is
\begin{equation}
    f = 2 \Omega \sin(\theta) \ ,
\end{equation}
for a spherical body. A hundred years after the work of Laplace, Lord Kelvin used this model (free of tidal forcing) to compute the normal-mode oscillations of a local plane, assuming a constant Coriolis parameter $f$ \cite{thomson18801}. He exhibited different kinds of plane-wave solutions, namely the \textit{surface gravity (Poincaré)} modes, the \textit{geostrophic} modes (for which the flow is stationary as the pressure gradient is exactly balanced by the Coriolis force) and \textit{coastal Kelvin} modes which exist only in the presence of a boundary and propagate in one direction. However, this \textit{$f$-plane approximation} is not relevant at the equator, where the parameter $f$ vanishes, thus virtually cancelling the Coriolis effects within the frame of this approximation. Rossby bypassed this problem by introducing the so-called \textit{$\beta$-plane approximation}, which amounts to assuming locally linear variation of the Coriolis parameter $f$ with latitude \cite{rossby1939relation}. This simple approximation led to qualitatively accurate understanding of many previously observed phenomena in geophysical fluid dynamics, such as the western intensification of wind-driven currents \cite{stommel1948westward,munk1950wind}, the oscillations of mid-latitude jets \cite{rossby1948displacements} or the equatorial trapping of gravity waves. Using this approximation at the equator with constant $\beta = \dd f / \dd y$, where $y$ is the meridional distance from the equator, Matsuno computed the spectrum of equatorial waves of the shallow-water model \cite{matsuno1966}. This spectrum has a discrete set of low-frequency \textit{planetary (Rossby)} waves and \textit{inertia-gravity} waves, plus two branches of modes transiting from the first to the second as the zonal wave number $k$ goes from negative to positive values, namely the \textit{Yanai} (or \textit{mixed-Rossby-gravity}) and \textit{equatorial Kelvin} modes.\\

\begin{figure}[H]
    \centering
    \includegraphics[scale=0.53]{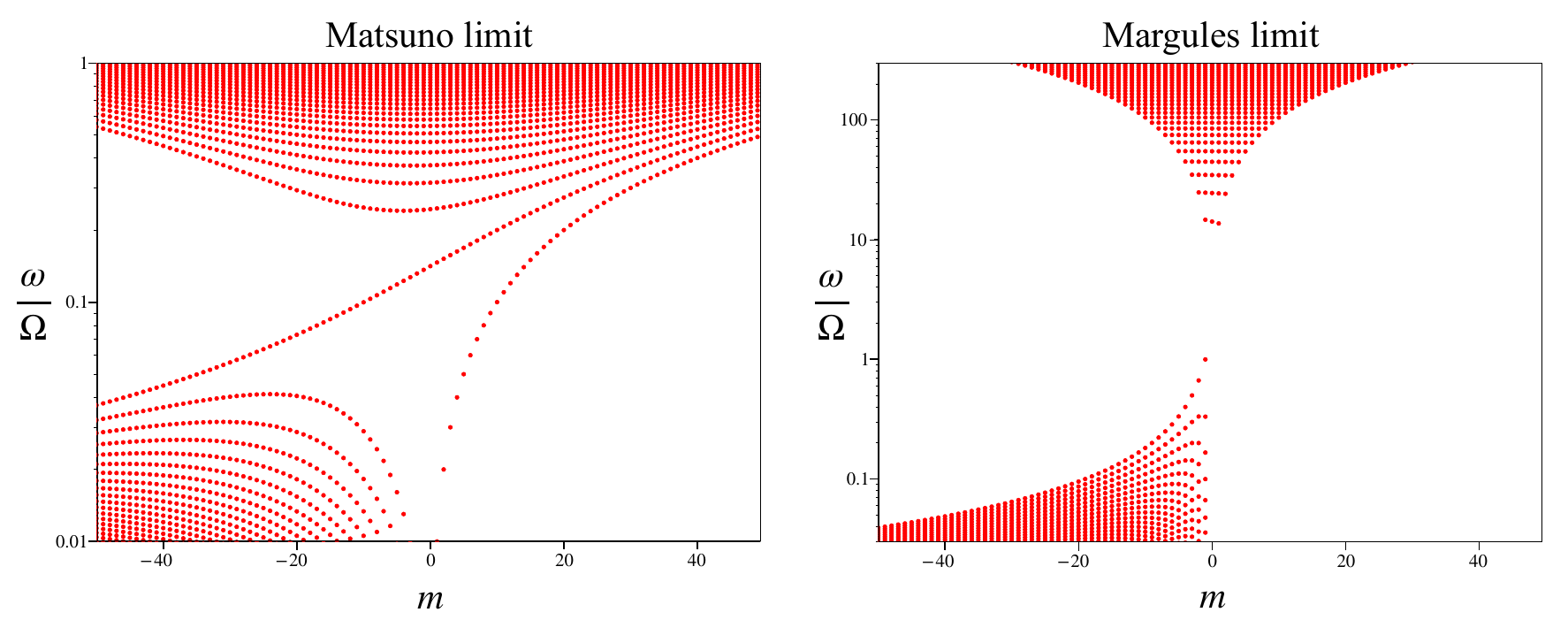}
    \caption{\label{fig:asymptotic_regimes} Dimensionless frequencies of the shallow-water model on the rotating sphere for two values of the parameter $\epsilon = c/\Omega R$ (left: $\epsilon = 0.01$, strong rotation; right: $\epsilon = 10$, weak rotation), in function of the azimuthal wave number $m \in \mathbb{Z}$, computed with \texttt{Dedalus} \cite{vasil2019tensor} (see Appendix \ref{apx:numerics} for numerical methods). In the Matsuno limit (left panel, $\epsilon \ll 1$), two distinct branches of frequencies transit through the gap between Rossby and inertia-gravity waves, whereas in the Margules limit (right panel, $\epsilon \gg 1$) these are clearly separated by an unfilled frequency gap. Note that the frequency axes in this figure are in logarithmic scale, as well as in Figure \ref{fig:transition}, in order to visually appreciate the transition between the Matsuno and Margules regimes. All the other frequency plots in the article are displayed in linear scale.}
\end{figure}

However, the equatorial $\beta$-plane approximation used by \cite{matsuno1966,delplace2017} (and many other studies of geophysical fluid dynamics) is limited in the sense that it only accounts for the curvature of the surface in the variation of the Coriolis parameter, at linear order, and not in the metric. It thus ignores the existence of the poles, and generates a spectrum of solutions defined on the unbounded domain across the equator, which seems paradoxical. Nevertheless, the solutions are accurate as long as they remain trapped around the equator, in the so-called \textit{Yoshida waveguide} \cite{yoshida1959theory,matsuno1966}, where the linear approximation holds. The trapping length, called the \textit{equatorial radius of deformation}, is given by
\begin{equation} \label{eq:equatorial_radius}
    L_{\rm eq} = \sqrt{\frac{c}{\beta}} \ ,
\end{equation}
where $c$ is the speed of surface waves without rotation. $L_{\rm eq}$ thus needs to be much smaller than $R$, the radius of the planet. This condition is satisfied for fast-rotating planets, compared to the timescale of propagation of the waves at their surface. For the first baroclinic mode (i.e. the fastest one) in the equatorial ocean on Earth, the equatorial radius of deformation is approximately $300$ kilometers (see e.g. \cite{vallis2017atmospheric}, p. 304), so the $\beta$-plane approximation is quite accurate in this context. However, when it comes to global oscillations of larger scales, both the curved metric and the actual sine variation of $f$ with latitude must be taken into account. Moreover, the quantisation of the zonal wave number $k$ cannot be ignored, especially when the wavelength is not small compared to the radius $R$. The full spectrum of shallow-water waves on the sphere is a peculiar problem, as the sphere is an unbounded but finite domain. The eigenvalue problem has no exact solution in the general case, however it has been extensively investigated through a variety of analytic approaches (e.g. \cite{margules1980air,hough1898v,longuet1968,bridger1980long,muller1995shallow,dellar2011variations,paldor2015}), and the frequencies and wave functions can be well-approximated with analytical expressions in the different asymptotic regimes. In particular, in the regime $L_{\rm eq} \ll R$ (hereafter referred to as \textit{Matsuno limit}), the spectrum of equatorial shallow-water waves is well-approximated by Matsuno's spectrum and thus clearly exhibits two branches of modes with eastward group velocity transiting through the frequency gap between the different \textit{wavebands} (in the rest of the paper, we will refer to the different groups of shallow-water modes as wavebands. There are three wavebands in the shallow-water model: the Rossby waveband and two equivalent inertia-gravity wavebands). In contrast, those seem to be absent from the frequency spectrum in the other limit (see Figure \ref{fig:asymptotic_regimes}), i.e. when $\Omega$ is small compared to $c/R$, hereafter referred to as \textit{Margules limit} \cite{margules1980air}. One would think that these branches somehow disappear as $L_{\rm eq}$ and $R$ become comparable, however there is no analytical solution of the eigenvalue problem in this intermediate situation to show exactly how.

Alternatively, a few years ago, the presence of these two transiting branches in the spectrum of shallow-water waves on the unbounded $\beta$-plane was interpreted as a manifestation of an underlying topological property. Indeed, after noticing the strong resemblance of this $+2$ \textit{spectral flow} of modes with the topologically-protected modes crossing the gap in certain insulating materials, \cite{delplace2017} applied the same arguments used in condensed matter physics to define topological integers associated to this spectral flow, the \textit{Chern numbers} or \textit{topological charges}. These integers characterise the phase singularities of Kelvin's plane-wave solutions, which are computed for constant $f$ and thus require only to diagonalise a 3-by-3 matrix. Nevertheless, they happen to be equal to the number of modes gained by the associated \textit{wavebands} in the more elaborated $\beta$-plane model. This counter-intuitive result, which connects the topological properties of Kelvin's $f$-plane modes to a spectral property of Matsuno's $\beta$-plane spectrum, is a consequence of the more general \textit{index theorem} \cite{faure2019manifestation,delplace2022berry,QinFu22}. It has been applied to predict the conditions of existence and number of modes transiting across the frequency gap in a variety models from a great variety of domains \cite{perrot2019topological,venaille2021,perez2022unidirectional,leclerc2022topological,parker2020topological,qin2023topological,wang2009observation,souslov2017topological,shankar2017topological,nash2015topological,khanikaev2015topologically}.\\

This paper intends to shed a new light on the loss of gapless Kelvin and Yanai modes on the sphere, not with exact analytical resolution but by combining the study of the structure of modes with spectral topology. This analysis relies on numerous previous works in which topology brought new insights on wave properties in geophysical and astrophysical media \cite{delplace2017,perrot2019topological,venaille2021,perez2022unidirectional,leclerc2022topological,perez2022topological}. In the first part, we introduce the equations of the rotating shallow-water model on the sphere, and the relevant dimensionless parameters that characterise it. We recall the asymptotic solutions in both Matsuno and Margules limits. In the second part, we discuss the concept of \textit{modal flow} and show that the latter is equal to zero for the spectrum of shallow-water waves on the rotating sphere, whereas it is equal to $+2$ on the unbounded $\beta$-plane. In the third part, we provide a topological interpretation of this result. We compute the Chern numbers, which are associated to the points where the symbol of the wave operator has multiple eigenvalues. We show that the metric term induces degeneracy points of non-zero Chern numbers, in addition to the one exhibited by \cite{delplace2017}. This analysis reconciles the bulk-interface correspondence established by Delplace \textit{et al} on the unbounded $\beta$-plane with the disappearance of the equatorial spectral flow in spherical geometry.

\section{The linearised shallow-water model on the rotating sphere} \label{sec:model}

\begin{figure}[H]
    \centering
    \includegraphics[scale=0.45]{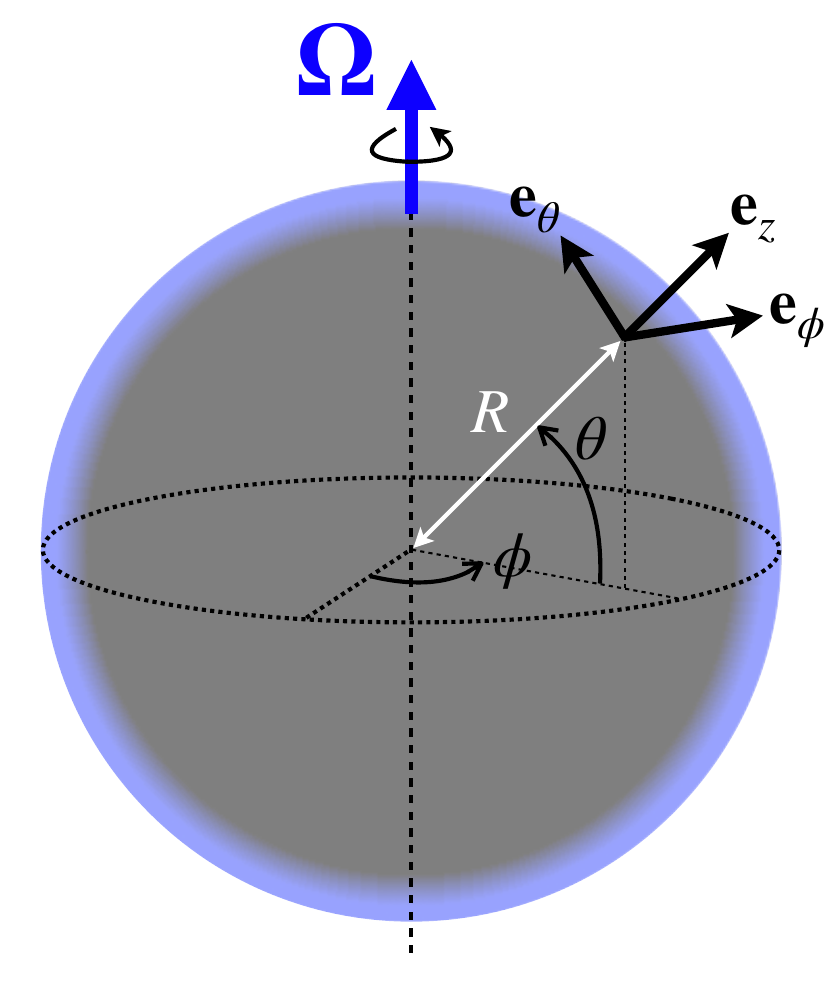}
    \caption{\label{fig:geometry_sphere} Geometry of the rotating shallow-water model on the sphere. The spherical coordinates and associated unit vectors are different than the usual ones, $\theta$ being the latitude instead of the colatitude.}
\end{figure}

\subsection{Linearised equations}

We consider the linearised shallow-water equations of an inviscid fluid layer on top of a sphere of radius $R$, rigidly rotating with constant rate $\mathbf{\Omega}$ (Figure \ref{fig:geometry_sphere}). The fluid is initially at rest in the rotating frame. Besides, assuming that the gravity $g$ at the surface is such that $g \gg R \Omega^2$, we ignore the centrifugal effects. Therefore, the rest state is that of a quiet layer of constant depth $h_0$. Noting the perturbed fields with $'$, we define $\mathbf{v} = \mathbf{v}'(\mathbf{x},t)$ and $h = h_0 + h'(\mathbf{x},t)$ the two-component velocity and height of the fluid, respectively, both functions of time $t$ and position $\mathbf{x}$ on the sphere. The linearised shallow-water equations can be conveniently expressed in terms of rescaled perturbation fields, i.e. $\Tilde{\mathbf{v}} = \sqrt{h_0} \ \mathbf{v}' = \Tilde{u} \, \mathbf{e}_\phi + \Tilde{v} \, \mathbf{e}_\theta$ and $\Tilde{h} = \sqrt{g} \ h'$, as
\begin{subequations} \label{eq:linearised_SW}
\begin{equation}
    \frac{\partial \Tilde{\mathbf{v}}}{\partial t} = -c \nabla \Tilde{h} - \mathbf{f} \times \Tilde{\mathbf{v}} \ ,
\end{equation}
\begin{equation}
    \frac{\partial \Tilde{h}}{\partial t} = -c \nabla \cdot \Tilde{\mathbf{v}} \ ,
\end{equation}
\end{subequations}
where $c = \sqrt{gh_0}$ is the constant phase speed and $\mathbf{f} = (2 \mathbf{\Omega} \cdot \mathbf{e}_z) \mathbf{e}_z$ is the projection of $2 \mathbf{\Omega}$ on the local unit vector $\mathbf{e}_z$ normal to the surface (traditional Coriolis parameter). Since none of the parameters appearing in Equations \eqref{eq:linearised_SW} depend on time or longitude, the rotating sphere acts as a waveguide trapping zonally-propagating waves in the meridional direction. We can thus expand any perturbation on the set of Fourier modes $X(\theta) \ee^{\ii(m \phi - \omega t)}$, where $X$ is any of the three dependent variables of Equations \eqref{eq:linearised_SW}, $(\phi,\theta)$ the longitudinal and latitudinal coordinates, respectively, and $(m,\omega)$ the azimuthal wave number and frequency associated with the Fourier mode, which corresponds to the solutions of
\begin{subequations} \label{eq:Fourier_SW}
\begin{equation} \label{eq:Fourier_v'}
    \omega \Tilde{\mathbf{v}} = -\ii c \nabla \Tilde{h} - \ii \mathbf{f} \times \Tilde{\mathbf{v}} \ ,
\end{equation}
\begin{equation} \label{eq:Fourier_h'}
    \omega \Tilde{h} = - \ii c \nabla \cdot \Tilde{\mathbf{v}} \ .
\end{equation}
\end{subequations}

The core of the discussion of this article is the metric term that arises from the divergence operator $\nabla \cdot$ in Equation \eqref{eq:Fourier_h'}, owing to the geometrical curvature of the surface. This point will be discussed in section \ref{sec:topology}.

\subsection{Lamb parameter and asymptotic solutions} \label{part:asymptotic_solutions}

If one rather considers the dimensionless frequency $\omega / \Omega$, the eigenvalue problem \eqref{eq:Fourier_SW} depends only on one dimensionless parameter:
\begin{equation}
    \epsilon = \frac{c}{R \Omega} \ .
\end{equation}

The quantity $2/\epsilon$ is sometimes called the \textit{Lamb parameter} \cite{muller1995shallow}, not to be confused with the Lamb frequency used in asteroseismology \cite{aerts2010asteroseismology}. $1/\epsilon$ gives the traversal time of a gravity wave over the spherical body in planetary days. Equations \eqref{eq:Fourier_SW} in dimensionless units are
\begin{subequations} \label{eq:Fourier_SW_bis}
\begin{equation} \label{eq:Fourier_v'_bis}
    (\omega / \Omega) \Tilde{\mathbf{v}} = -\ii \epsilon \Tilde{\nabla} \Tilde{h} - 2 \ii \sin (\theta) \mathbf{e}_z \times \Tilde{\mathbf{v}} \ ,
\end{equation}
\begin{equation} \label{eq:Fourier_h'_bis}
    (\omega / \Omega) \Tilde{h} = - \ii \epsilon \Tilde{\nabla} \cdot \Tilde{\mathbf{v}} \ ,
\end{equation}
\end{subequations}
where $\Tilde{\nabla} = R \nabla$ is the gradient operator on the unit sphere. The parameter $\epsilon$ naturally appears in asymptotic expansions to compute the approximate spectrum of shallow-water waves \cite{muller1995shallow,paldor2015} or their ray paths \cite{longuet1965planetary} on the sphere. One can expect to asymptotically recover the Matsuno spectrum as $\epsilon \rightarrow 0$. Indeed, Expression \eqref{eq:equatorial_radius} yields $L_{\rm eq} / R = \sqrt{\epsilon / 2}$ for the equatorial radius of deformation, which means that, for small $\epsilon$, the wave functions are equatorially trapped with angular spreading of order $\sqrt{\epsilon}$, thus recovering the Cartesian $\beta$-plane approximation made by Matsuno. In this limit, the solutions of \eqref{eq:Fourier_SW} are appropriately described with Hermite polynomials, which are exact solutions on the unbounded $\beta$-plane (see \ref{part:Matsuno}).\\

\begin{figure}[H]
    \centering
    \includegraphics[scale=0.5]{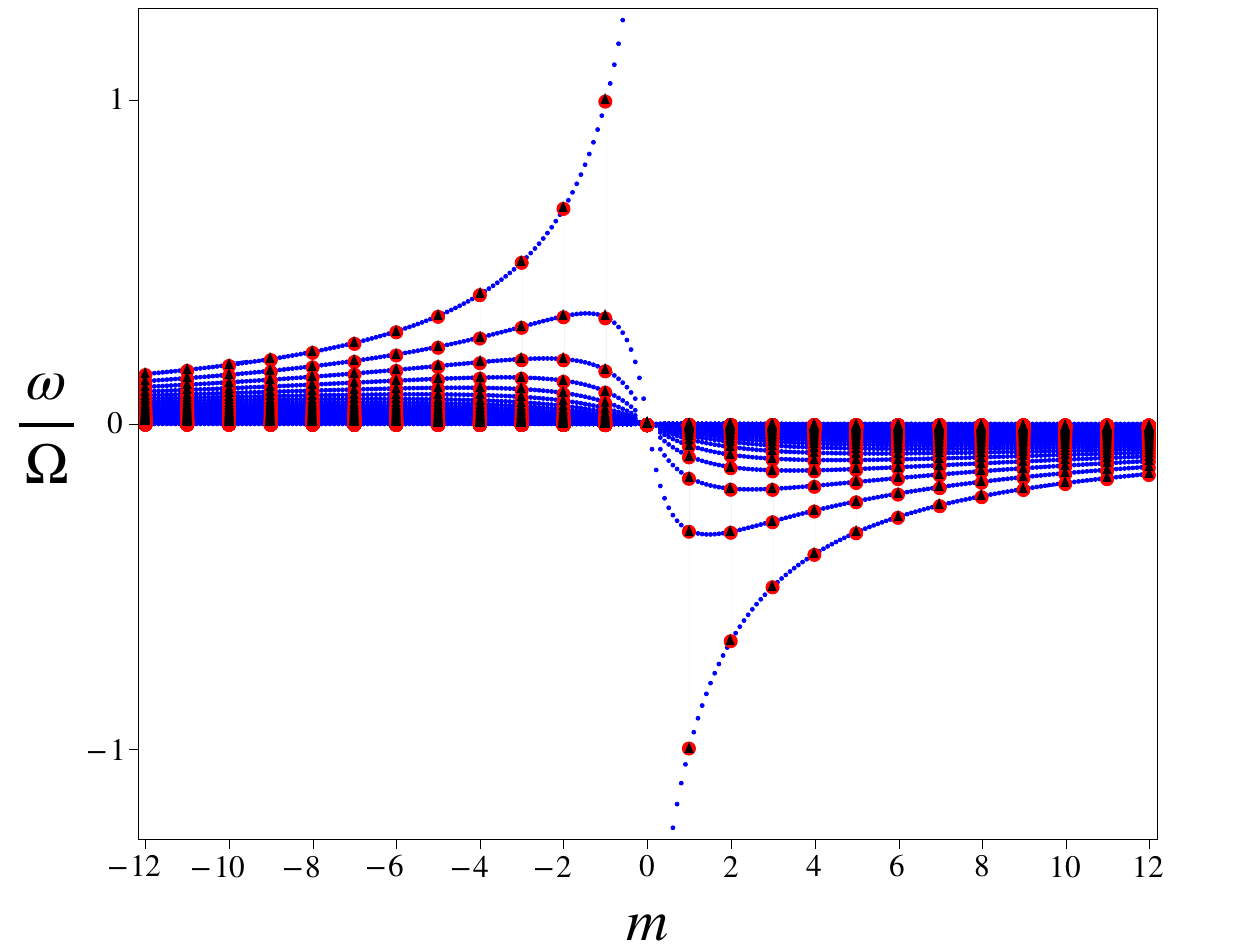}
    \caption{\label{fig:Rossby_spherical} Comparison between Expression \eqref{eq:Rossby_sphere} (black triangles) and numerically-computed Rossby frequencies for $\epsilon = 5$. The red dots are obtained by projecting Equations \eqref{eq:Fourier_SW} on \texttt{Dedalus}' curvilinear spectral basis \cite{vasil2019tensor}, whereas the blue dots are computed from the problem \eqref{eq:Schrodinger_SW} on $(-\pi/2,\pi/2)$, not necessarily with integer values of $m$, projecting the wave functions on the basis of Chebyshev polynomials \cite{burns2020} (see Appendix \ref{apx:numerics}). The accuracy between the numerically-converged frequencies for $\epsilon=5$ and Margules' asymptotic formula \eqref{eq:Rossby_sphere} is approximately $0.4 \%$.}
\end{figure}

Conversely, for larger values of $\epsilon$, the solutions are expected to spread away from the equator and eventually over the whole sphere. In the limit $\epsilon \gg 1$, sometimes called the \textit{Margules limit} \cite{muller1995shallow}, there are two kinds of eigenvalues, originally found by Margules \cite{margules1980air} and Hough \cite{hough1898v}, that we recall here:

\begin{itemize}
    \item The ones such that $\omega = \mathcal{O}(c/R)$. One recovers the non-rotating shallow-water equations from \eqref{eq:Fourier_SW}. The wave functions are those of the Laplacian operator on a prolate spheroid, i.e. the prolate spheroidal wave functions \cite{zaqarashvili2009}, and we have $R \omega / c \simeq \sqrt{\ell (\ell + 1)} - m / \epsilon$ (see e.g. \cite{muller1994,muller1995shallow}) with non-zero integers $\ell$ and $|m| \leq \ell$. These are global surface gravity oscillations on the sphere, perturbed by a small Coriolis term. For these solutions, the number of zeros of $\Tilde{v}$ on the open interval $(-\pi/2 , \pi/2)$ is given by $p = \ell - |m| + 1$ \cite{muller1995shallow}.
    
    \item The ones such that $\omega = \mathcal{O}(\Omega)$. These are quasi-geostrophic planetary (Rossby) modes. From \eqref{eq:Fourier_SW}, it is straightforward to check that they obey $|\Tilde{\nabla} \cdot \mathbf{v}'| = \mathcal{O} \left( |\mathbf{v}'| / \epsilon^2 \right)$, i.e. that these modes have asymptotically divergence-free velocity. Eliminating the other variables from Equations \eqref{eq:Fourier_v'}, we end up with a second-order differential equation for $V = \cos^\frac{3}{2} (\theta) \, \Tilde{v}$:
    \begin{equation} \label{eq:Schro_v}
        -\frac{\dd^2 V}{\dd \theta^2} + \left[ \left( m^2 - \frac{1}{4} \right) \tan^2 (\theta) + \left( m^2 - \frac{1}{2} + \frac{2m \Omega}{\omega} \right) \right] V = 0 \ .
    \end{equation}
    
    Solutions of Equation \eqref{eq:Schro_v} are provided for instance by \cite{tacseli2003exact}. They yield
    \begin{equation} \label{eq:Rossby_sphere}
        \frac{\omega}{\Omega} = \frac{-2m}{m^2 + (2p + 1)|m| + p(p + 1)} \quad (\text{with} \quad p = 0,1,2...) \ ,
    \end{equation}
    where the index $p$ indicates the number of zeros of the meridional velocity $\Tilde{v}$ in the open interval $(-\pi/2 , \pi/2)$. In the Margules limit, the largest Rossby frequency is thus $\omega = \Omega$, for $m = -1$ and $p=0$. The Rossby modes with $p=0$ constitute the westward Yanai modes \cite{paldor2018mixed}. Expression \eqref{eq:Rossby_sphere}, which is exact for any azimuthal wave number $m$ in the Margules limit $\epsilon \gg 1$, is the same as (3.2c) obtained by \cite{muller1995shallow} through a covariant formulation of the wave equations \eqref{eq:linearised_SW}. Figure \ref{fig:Rossby_spherical} provides a comparison of Expression \eqref{eq:Rossby_sphere} with numerical simulations, demonstrating the convergence of the Rossby waveband toward the Margules limit as $\epsilon$ increases.
\end{itemize}

There is a large frequency gap between these two wavebands, in which the Kelvin and Yanai modes of the Matsuno spectrum are absent (see Figure \ref{fig:asymptotic_regimes}).

\subsection{Matsuno and Margules limits on planets and stars}

As far as the shallow-water model is valid, the Matsuno regime is common to most fast-rotating planets, but also stars like brown dwarfs \cite{tan2020}. Observations of the $\delta$ Scuti star Rasalhague allows one to estimate $\epsilon \simeq 0.36$ \cite{monnier2010}. Atmospheres of exoplanets cover a large range of values of $\epsilon$ (see e.g. \cite{showman2010atmospheric}, Table 1), as well as the atmosphere of Earth, on which there is a variety of values of $\epsilon$ depending on the type of waves considered \cite{muller1995shallow,shamir2023matsuno}. For instance, the measurements of Kelvin and Rossby-Haurwitz waves in the atmosphere of Venus presented by \cite{del1990} lead to the estimation $\epsilon \simeq 0.9$, whereas $\epsilon \approx 70$ for the atmospheric waves considered by \cite{showman2010atmospheric}. On Earth, equatorial Kelvin waves have been detected both in the equatorial ocean \cite{johnson1993structure,sprintall2000semiannual} and atmosphere \cite{kiladis2009}, and their implication in a variety of geophysical phenomena (e.g. the El Ni\~no event and the Madden-Julian Oscillation) is well understood. The first baroclinic modes (the fastest ones) are such that $c \simeq 2 \; {\rm m \; s}^{-1}$ in the equatorial ocean, thus $\epsilon \simeq 4.3 \; 10^{-3}$, and $c \simeq 25 \; {\rm m \; s}^{-1}$ in the atmosphere, thus $\epsilon \simeq 5.4 \; 10^{-2}$ (see \cite{vallis2017atmospheric}, p. 304). However, the barotropic modes in the terrestrial atmosphere propagate at phase speed $c \simeq 300 \; {\rm m \; s}^{-1}$, which corresponds to $\epsilon \simeq 0.65$. This value is consistent with the data studied by \cite{sakazaki2020array}. Surface motions on giant planets of our solar system are the subject of many recent works \cite{menou2009atmospheric,showman2010atmospheric,showman2018global,gavriel2021number}. Atmospheric waves on Jupiter and Saturn are mostly in the Matsuno regime. For instance, eastward-propagating motions strongly localised at the equator on Jupiter have been interpreted as Kelvin waves \cite{legarreta2016}, which is consistent with the estimation $\epsilon \simeq 1.4 \; 10^{-4}$.

\subsection{Opening of a spectral gap as \texorpdfstring{$\boldsymbol{\epsilon}$}{} increases}

\begin{figure}[H]
    \centering
    \includegraphics[scale=0.54]{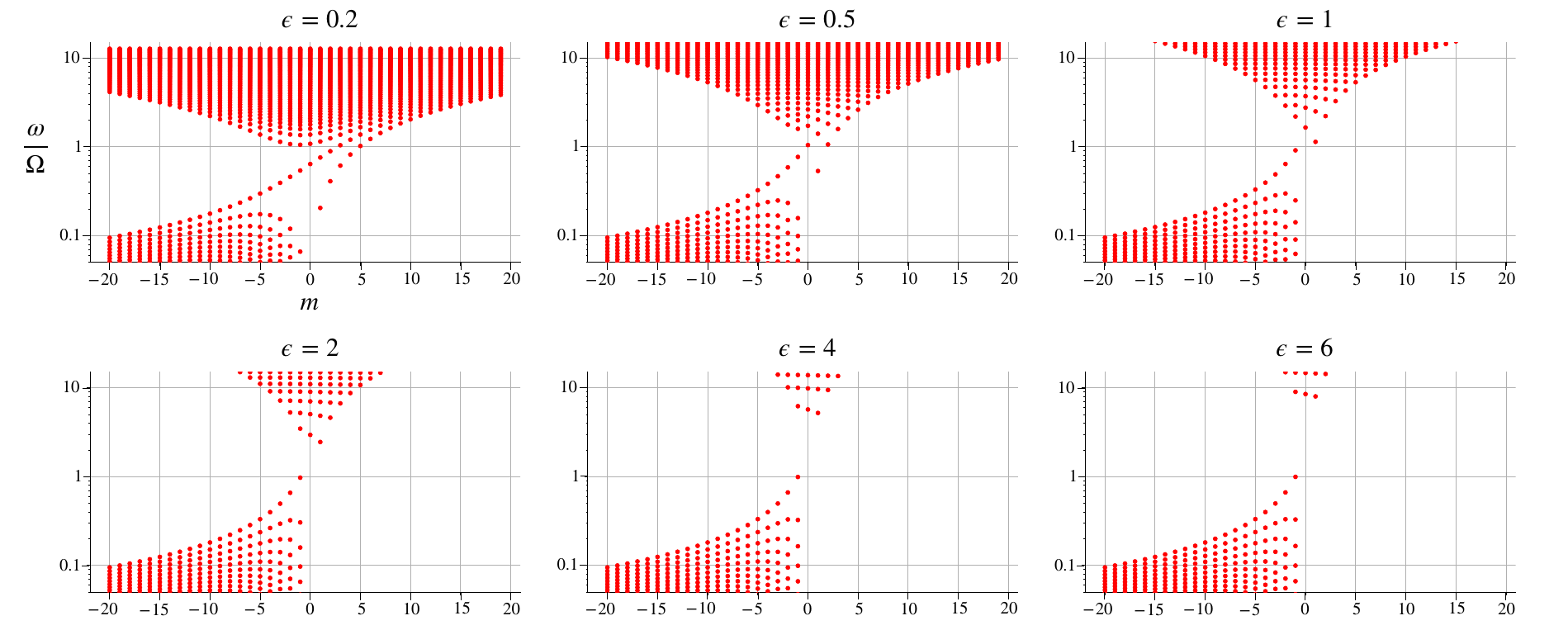}
    \caption{\label{fig:transition} Numerically-computed frequencies $\omega$ for increasing values of $\epsilon$ and $|m| \leq 20$. These converged values are obtained by projecting Equations \eqref{eq:Fourier_SW} on a basis of spin-weighted harmonics, a spectral method implemented in the \texttt{Dedalus} solver \cite{vasil2019tensor} (see Appendix \ref{apx:numerics}). While for small $\epsilon$ two distinguishable spectral branches, the Yanai and Kelvin modes, transit through the gap between Rossby and gravity modes, these branches progressively break down and a clear gap opens up, separating Rossby and gravity frequencies.}
\end{figure}

In the Margules limit ($\epsilon \gg 1$), the minimal inertia-gravity frequency is equal to $\sqrt{2} \: c/R$ and the maximum Rossby frequency is $\Omega$, which yields a frequency gap of width $\Omega \left( \sqrt{2} \: \epsilon - 1 \right)$ in which there is no mode. Conversely, in the Matsuno limit ($\epsilon \ll 1$), the frequency gap between Rossby and inertia-gravity modes is filled with the Yanai and Kelvin modes. However, talking about a frequency gap and branches of modes is abusive in this situation. Indeed, contrary to the zonal wave number $k$ of zonally-propagating waves on the unbounded $\beta$-plane, the azimuthal wave number $m$ is not a continuous parameter, since it only takes discrete integer values. Therefore the spectrum $(m,\omega)$ does not consist of continuous branches but rather a discrete set of points, which means that, strictly speaking, there are frequency gaps between any allowed values of $\omega$. Nevertheless, the equatorial Yanai and Kelvin modes define a clear connection between the Rossby and inertia-gravity wavebands on the unbounded $\beta$-plane, which is still visible in the shallow-water spectrum on the sphere with small $\epsilon$, even if $m$ takes discrete values (see Figures \ref{fig:asymptotic_regimes} and \ref{fig:transition}). Such a connection in the spectrum is referred to as spectral flow, which is usually defined for a continuous \textit{spectral parameter} such as $k$ in unbounded flow models \cite{delplace2017,shankar2017topological,perrot2019topological,parker2020topological,venaille2021,perez2022unidirectional,leclerc2022topological,zhu2023topology}. The purpose of the following section is to extend the concept of spectral flow to the present situation, i.e. with a discrete wave number $m$, and show that the net number of modes gained by the inertia-gravity waveband, when those modes are naturally labelled by the zeros of their meridional velocity, is actually zero for any value of $\epsilon$, even if there are transiting frequencies for small $\epsilon$ (see Figure \ref{fig:transition}).

\section{Modal flow and zeros of the meridional velocity} \label{sec:zeros}

To this day, Equations \eqref{eq:Fourier_SW} do not have an exact solution for arbitrary values of $\epsilon$, although good approximations have been found in certain limits, especially in short wavelength ranges \cite{longuet1968,paldor2015}, or for special values of $\omega$ and $\epsilon$ \cite{muller1995shallow}. In this study, we are interested in the transition between the Matsuno and Margules limits, i.e. the evolution of modes and frequencies for finite values of $\epsilon$. For this reason, the analysis of this section is mostly based on numerical integration of Equations \eqref{eq:Fourier_SW}, with the spectral solver \texttt{Dedalus} \cite{vasil2019tensor}. The point of this section is the following: the Matsuno spectrum, i.e. the frequencies $\omega$ of zonally-propagating equatorial waves on the unbounded $\beta$-plane, consists of a discrete set of continuous functions of the zonal wave number $k$, or \textit{spectral branches}. These branches can be indexed by the number of zeros of the meridional velocity $\Tilde{v}(\theta)$. For shallow-water waves on the sphere, however, the modes and frequencies depart from Matsuno's solutions as $\epsilon$ increases, wave functions spread away from the equator and new zeros of $\Tilde{v}(\theta)$ appear at non-zero latitudes, modifying the natural order of mode branches established by Matsuno. This affects the conclusion regarding the very concept of spectral flow in this model, which must be replaced by a more accurate notion that we name the modal flow.

\subsection{Reminder: the Matsuno spectrum and wave functions} \label{part:Matsuno}

\begin{figure}[H]
    \centering
    \includegraphics[scale=0.42]{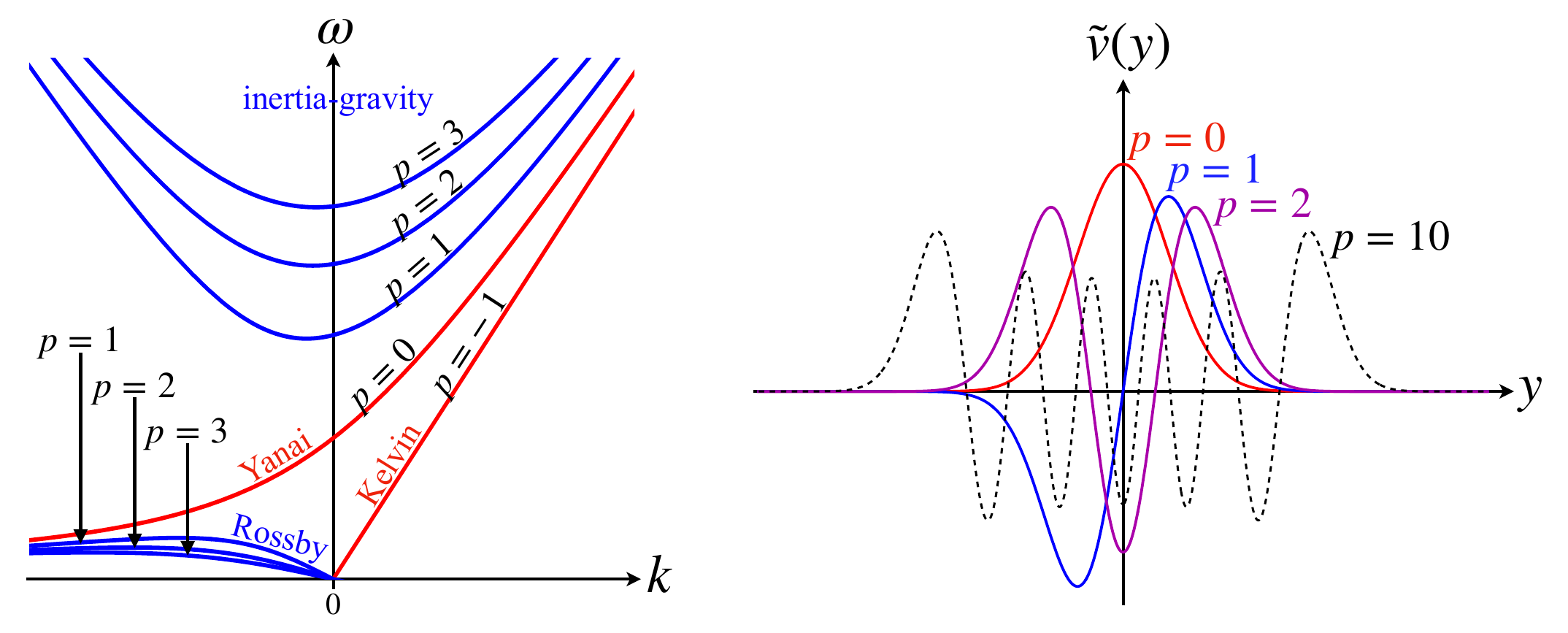}
    \caption{\label{fig:Matsuno} The Matsuno spectrum and wave functions for equatorial shallow-water waves on the unbounded $\beta$-plane \cite{matsuno1966}. Left: plot of the dispersion relation \eqref{eq:Matsuno_dispersion}, showing the branches $-1 \leq p \leq 3$ for positive frequencies. The Yanai and Kelvin branches transit across the frequency gap between the Rossby and inertia-gravity wavebands, as $k$ increases. Right: Amplitude of the meridional velocity in the $y$ direction, given by Expression \eqref{eq:Matsuno_Hermite}. The branch index $p$ is equal to the number of zeros of $\Tilde{v}(y)$.}
\end{figure}

The shallow-water equations for equatorial waves on the unbounded $\beta$-plane are the same as \eqref{eq:Fourier_SW}, only with Cartesian coordinates instead of the spherical ones ($x$ is the zonal coordinate pointing eastward and $y$ is the meridional coordinate pointing northward, with $y=0$ defining the equator), and $f=\beta y$. Once again, we focus on plane waves propagating in the zonal direction, i.e. solutions in the form $X(y) \ee^{\ii(k x - \omega t)}$, where the wave number $k$ can take continuous values. As explained in the introduction, this model captures the physics of equatorial waves in the limit of small $\epsilon$. The solutions found by \cite{matsuno1966} obey the dispersion relation
\begin{equation} \label{eq:Matsuno_dispersion}
    \frac{\omega^2}{c^2} - k^2 - \frac{\beta k}{\omega} - \frac{\beta}{c} (2p+1) = 0 \quad (\text{with} \quad p = -1,0,1,2...) \ ,
\end{equation}
and the corresponding wave functions for the meridional velocity are given by
\begin{equation} \label{eq:Matsuno_Hermite}
    \begin{split}
        \Tilde{v}(y) &\propto H_p \left( y \sqrt{\frac{\beta}{c}} \right) \ee^{-\frac{\beta}{2c} y^2} \quad \text{for} \quad p \geq 0 \ , \\
        \Tilde{v}(y) &= 0 \quad \text{for} \quad p=-1 \ ,
    \end{split}
\end{equation}
with the Hermite polynomials $H_p$. As depicted in Figure \ref{fig:Matsuno}, the dispersion relation \eqref{eq:Matsuno_dispersion} can be represented by a discrete set of continuous branches in $(k,\omega)$, each indexed by the integer $p$. For $p \geq 0$, this index is equal to the number of zeros of $\Tilde{v}$ in the meridional direction $y$. Branches with $p\  \geq 1$ are the Rossby and inertia-gravity modes. The branches $p=0$ and $p=-1$ are those of the Yanai and Kelvin modes, respectively. The Kelvin modes are purely zonal, i.e. with $\Tilde{v} = 0$. The dispersion relation \eqref{eq:Matsuno_dispersion} formally associates Kelvin modes with the index $p=-1$ and, even though it does not make sense in terms of number of zeros of $\Tilde{v}(y)$, it will be convenient to keep this index $-1$ in mind. It is also worth noticing that, for $p=0$, Equation \eqref{eq:Matsuno_dispersion} admits an additional solution $\omega = -ck$. This solution corresponds to a diverging mode on the unbounded $\beta$-plane, thus it is discarded as it is not physically acceptable.

\subsection{Zeros of the meridional velocity}

\begin{figure}[H]
    \centering
    \includegraphics[scale=0.55]{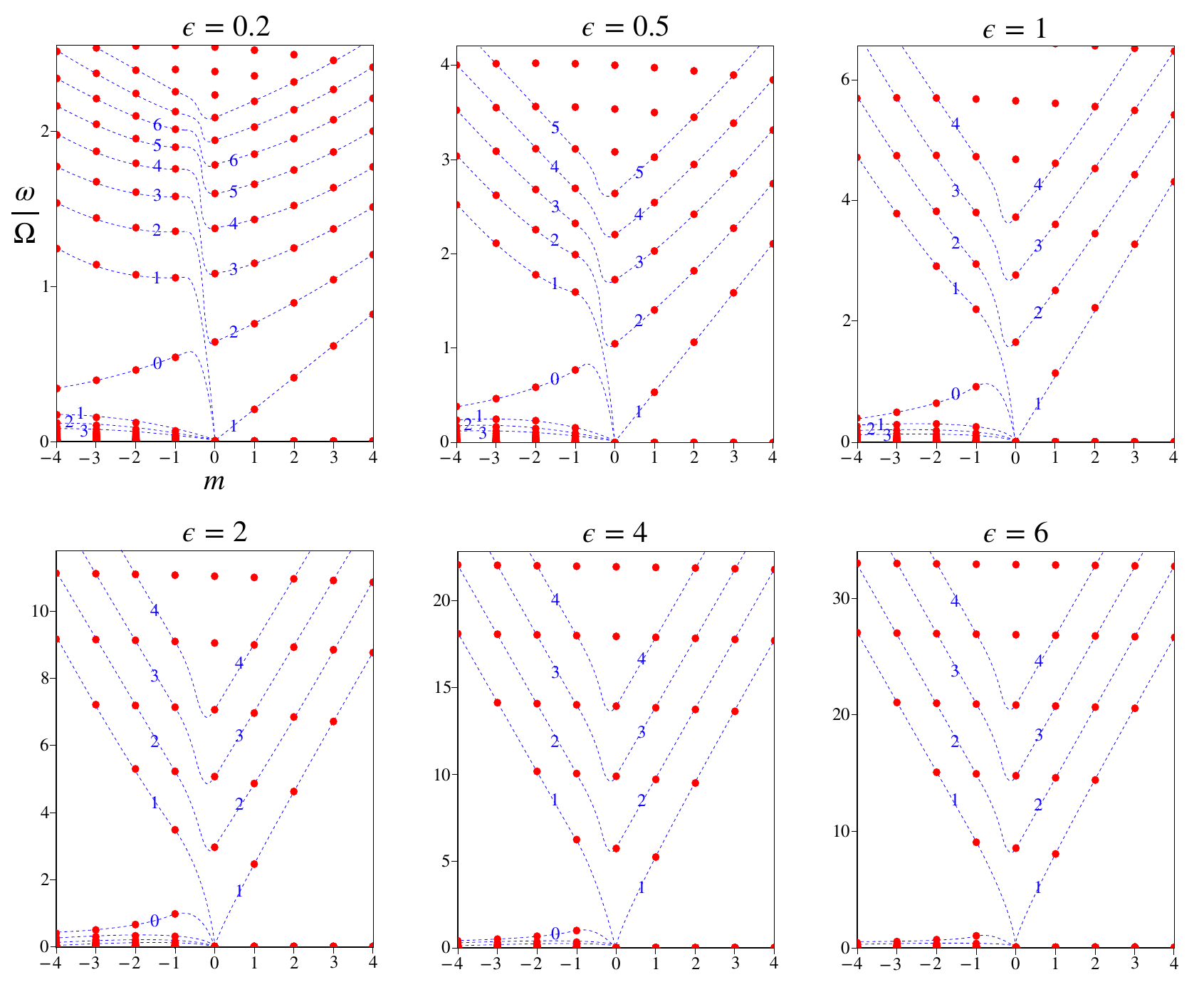}
    \caption{\label{fig:branches_v_label} Numerically-calculated frequencies of Equations \eqref{eq:Fourier_SW} (red dots). The modes with same number of zeros of $\Tilde{v}(\theta)$ are connected by dashed blue curves, which have no physical meaning. The number of zeros is indicated on the curves. For small $\epsilon$, the modal branches display an abrupt jump right before $m=0$. As $\epsilon$ increases, the spectrum progressively loses its east-west asymmetry and the spectral flow collapses.}
\end{figure}

The $+2$ spectral flow studied by \cite{delplace2017} corresponds to the Yanai and Kelvin modes, which transit from the Rossby waveband to the inertia-gravity waveband as $k$ increases. Beyond the fact that these modes form continuous curves in $(k,\omega)$, what identifies each of them is the number of zeros of the meridional velocity in the $y$ direction, as previously shown. Generally speaking, it is straightforward to define the spectral flow of an eigenvalue problem if there is a continuous parameter such as the zonal wave number $k$, because the spectrum thus consists of continuous branches. However, if the spectral parameter takes discrete values, one must find an alternative way to count the number of modes gained or lost by a waveband as this spectral parameter is swept. For shallow-water waves on the rotating sphere, the azimuthal wave number $m$ is a discrete parameter, nevertheless we will extrapolate the relation existing between the spectral flow of inertia-gravity waves and the number of zeros of $\Tilde{v}$ for the unbounded $\beta$-plane model. In other words, we will adopt the number of zeros of $\Tilde{v} (\theta)$ on the open interval $(-\pi/2 , +\pi/2)$ as an ordering parameter for the modes of the problem \eqref{eq:Fourier_SW} (see Figure \ref{fig:branches_v_label}). In fact, many works on the spectrum of shallow-water waves associate or \textit{label} modes according to the number of zeros of their meridional velocity \cite{hough1898v,longuet1968,iga1995transition,muller1995shallow,paldor2018mixed}. In the Matsuno spectrum, at negative $k$, there are modes with no zero of $\Tilde{v}$ in the Rossby waveband but not in the inertia-gravity waveband, whereas there are modes with no zero in the inertia-gravity waveband at positive $k$. Connecting them together forms the spectral flow of Yanai modes, which transits between the Rossby and inertia-gravity wavebands as $k$ increases. To clarify, in the rest of the paper, we will employ the term \textit{modal branch} to refer to modes with the same number of zeros of $\Tilde{v}(\theta)$ on the open interval $(-\pi/2 , \pi/2)$, as $m$ varies, and say that there is a \textit{modal flow} when such a modal branch transits between the Rossby and inertia-gravity wavebands. In contrast, the terms \textit{spectral branch} and \textit{spectral flow} will be used to refer to the continuous curves formed by the frequencies, which is a well-defined concept only if the spectral parameter is continuous, strictly speaking. Spectral and modal branches of the Matsuno spectrum are the same (see Figure \ref{fig:Matsuno}). In the shallow-water spectrum on the sphere, the spectral parameter $m$ is not continuous, yet one can argue that there is a spectral flow of frequencies in the Matsuno limit (see Figures \ref{fig:asymptotic_regimes} and \ref{fig:transition}), but not in the Margules limit.

\begin{figure}[H]
    \centering
    \includegraphics[scale=0.55]{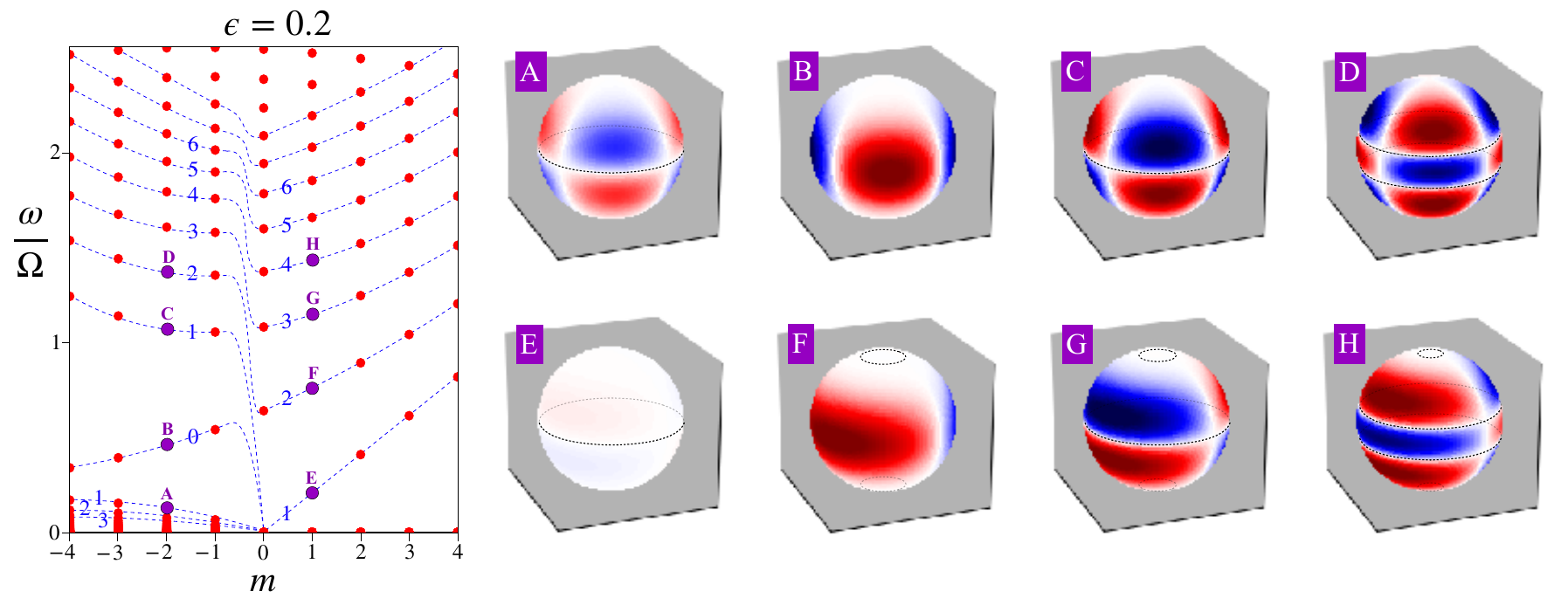}
    \caption{\label{fig:modes_0.2} Meridional velocity $\mathrm{Re} \left( \Tilde{v}(\theta) \ee^{\ii m \phi} \right)$ of modes for $\epsilon = 0.2$ and $m=-2,+1$, computed with \texttt{Dedalus} (red: positive values; blue: negative values). A is a Rossby mode, B and F are Yanai modes, E is a Kelvin mode, C, D, G and H are inertia-gravity modes. On the left, modes are connected by dashed blue curves according to number of zeros of $\Tilde{v}(\theta)$ on $(-\pi/2 , \pi/2)$. Modes F,G and H have zeros at high latitude, which are not visible on the plot since the wave functions vanish away from the equator (see Appendix \ref{apx:numerics}). Dashed black lines indicate the zeros of $\Tilde{v}(\theta)$, and the meridional lines of zeros correspond to the term $\ee^{\ii m \phi}$ with $m \neq 0$ for travelling waves.}
\end{figure}

\begin{figure}[H]
    \centering
    \includegraphics[scale=0.55]{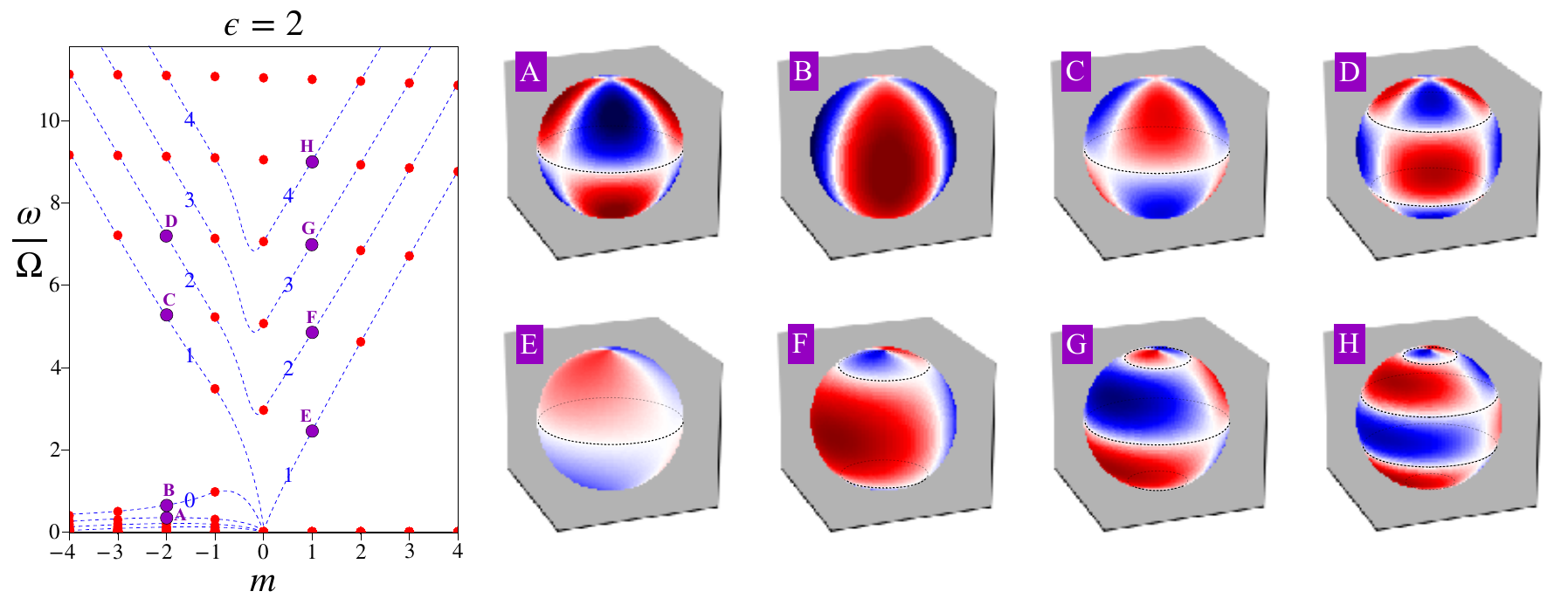}
    \caption{\label{fig:modes_2}Same modes as Figure \ref{fig:modes_0.2}, for $\epsilon = 2$. The wave functions are more spread in latitude and extend up to the poles, clearly revealing zeros at high latitudes. The meridional velocity of a Yanai mode has no zero for negative $m$ (B), but changes sign at two opposite latitudes for positive $m$ (F), contrary to the Yanai mode computed by Matsuno on the equatorial $\beta$-plane. Generally speaking, the meridional velocity of modes with eastward phase speed has two more zeros, compared to Matsuno's wave functions. Similarly, Kelvin modes (E) have zero meridional velocity on the unbounded $\beta$-plane, which is not true on the sphere. Note that the north-south symmetry of the wave functions is preserved on the sphere, which implies the parity of these additional zeros. This point is discussed in Appendix \ref{apx:symmetry}.}
\end{figure}

As shown in the previous section, for small $\epsilon$, the ratio between the equatorial radius of deformation and the sphere's radius is of order $\sqrt{\epsilon}$, which implies that the wave functions spread across the whole sphere even for moderate values of $\epsilon$, and thus the modes experience the discrepancy with the unbounded $\beta$-plane. The spreading of all the wave functions reveals additional zeros of $\Tilde{v}$ at non-zero latitude, which are virtually invisible when $\epsilon$ is too small and the modes are strongly trapped at the equator. In fact, all modes with positive phase speed (i.e. positive $m$, considering only the positive frequencies) have two additional zeros at opposite latitudes (see Figures \ref{fig:modes_0.2} and \ref{fig:modes_2}), compared to the same modes of the unbounded $\beta$-plane. However, the zeros of Rossby modes, which propagate westward, are unchanged. In other words, the index $p$ of the Matsuno modes propagating eastward is increased by $+2$ in spherical geometry. In particular, Yanai modes with negative $m$ have $p=0$ meridional zeros and $p=2$ zeros for positive $m$. Similarly, in contrast with the unbounded $\beta$-plane, Kelvin modes have a non-zero meridional velocity which cancels once at the equator, i.e. $p=1 = -1 + 2$. This $+2$ jump in number of zeros, which was already discussed by \cite{muller1995shallow}, is illustrated in Figure \ref{fig:branches_v_label}.

To conclude this part, we have shown that the net modal flow of shallow-water waves, equal to $+2$ on the unbounded $\beta$-plane, is null on the sphere. This is evident for large $\epsilon$ since the spectrum displays a strong east-west symmetry, i.e. the frequencies $\omega$ and wave functions are nearly the same for $m$ and $-m$. As $\epsilon$ becomes smaller, the spectrum loses this east-west symmetry, displays an apparent spectral flow of frequencies but the net modal flow remains zero. At small $\epsilon$, the spectrum resembles the Matsuno one as there are Kelvin and Yanai modes transiting across the gap for increasing wave number $m$. However, the Yanai modes do not form a modal branch because the eastward ones have two more zeros than the westward ones. 
In the present case, the nullity of this modal flow can be surprising regarding the bulk-interface correspondence established by \cite{delplace2017,tauber2019bulk,delplace2022berry}, which states that the number of transiting modes of the Matsuno spectrum is equal to a robust topological charge $+2$ at the equator. In the following we will show that the nullity of the net modal flow is actually in agreement with the index theorem, as the Chern numbers of the shallow-water spectrum on the sphere add up to zero, in contrast with the unbounded $\beta$-plane.

\section{Topology of the shallow-water model on the sphere} \label{sec:topology}

Identifying modes with the number of zeros of their $\Tilde{v}$ wave function, we showed that the net modal flow of shallow-water waves on the sphere is null, while it is $+2$ for the unbounded $\beta$-plane. Many studies \cite{faure2000topological,delplace2017,perrot2019topological,parker2020topological,venaille2021,fu2021topological,perez2022unidirectional,leclerc2022topological,delplace2022berry} showed that, when the frequencies $\omega$ can be expressed as the eigenvalues of a Hermitian differential wave operator $\mathcal{H}$, the number of modes gained by a waveband as the spectral parameter increases is equal to a topological invariant characterising the \textit{Weyl symbol} $H$ of the operator $\mathcal{H}$, in virtue of the \textit{index theorem} \cite{faure2019manifestation}. As demonstrated for instance by \cite{delplace2017,faure2019manifestation,delplace2022berry,venaille2023ray}, the presence of $+2$ transiting equatorial modes on the unbounded $\beta$-plane is ensured by a $+2$ Chern number characterising the degeneracy point of the symbol at $f=0$. More precisely, the Yanai and Kelvin branches constitute a footprint, in the spectrum of $\mathcal{H}$, of a degeneracy point of its symbol bearing a Chern number $+2$. In light of the discussion of section \ref{sec:zeros}, the aim of this section is to show that the previous conclusions can be inferred with the index theorem, as accounting for the spherical metric reveals new degeneracy points of the symbol of the wave operator, in addition to the unique degeneracy point of the problem on the unbounded $\beta$-plane. We will demonstrate that these additional degeneracy points bear topological charges which compensate the $+2$ Chern number of the equatorial degeneracy point, thus making the eigenbands of the symbol have zero total Chern numbers, which justifies the absence of transiting modes and the breaking of Yanai and Kelvin mode branches on the sphere for large $\epsilon$.

\subsection{Hermitian form, analogy with topographic shallow-water waves}

In order to apply the index theorem, we first need to express Equations \eqref{eq:Fourier_SW} in the form $\mathcal{H} X = \omega X$, where $\mathcal{H}$ is a Hermitian differential wave operator and the complex vector $X$ contains the three dependent variables $\Tilde{u}(\theta),\Tilde{v}(\theta)$ and $\Tilde{h}(\theta)$. To do this, we introduce the rescaled fields $u = \sqrt{\cos(\theta)} \, \Tilde{u}$, $v = \sqrt{\cos(\theta)} \, \Tilde{v}$ and $\eta = \sqrt{\cos(\theta)} \, \Tilde{h}$. Equations \eqref{eq:Fourier_SW} can thus be recast in the form of a matrix eigenvalue equation:
\begin{equation} \label{eq:Schrodinger_SW}
    \omega \begin{pmatrix} u \\[6pt] v \\[6pt] \eta \end{pmatrix} = \begin{pmatrix} 0 && \ii f(\theta) && \frac{c}{R} \frac{m}{\cos \theta} \\[6pt] -\ii f(\theta) && 0 && -\ii \frac{c}{R} \frac{\dd}{\dd \theta} - \ii \beta_{\rm g}(\theta) \\[6pt] \frac{c}{R} \frac{m}{\cos \theta} && -\ii \frac{c}{R} \frac{\dd}{\dd \theta} + \ii \beta_{\rm g}(\theta) && 0 \end{pmatrix} \begin{pmatrix} u \\[6pt] v \\[6pt] \eta \end{pmatrix} = \mathcal{H} \begin{pmatrix} u \\[6pt] v \\[6pt] \eta \end{pmatrix} \ ,
\end{equation}
with the Coriolis parameter $f = 2 \Omega \sin (\theta)$ and a \textit{metric $\beta$-term}
\begin{equation} \label{eq:metric_term}
    \beta_{\rm g} = \frac{c}{2 R} \tan (\theta) \ .
\end{equation}

The problem is that of identifying the eigenvalues of the wave operator $\mathcal{H}$, the 3-by-3 matrix of differential operators appearing in the RHS of Equation \eqref{eq:Schrodinger_SW}. It is a Hermitian matrix operator for the canonical scalar product of complex vector functions of $\theta$ (see Appendix \ref{apx:WW}). In consequence, unsurprisingly, the spectrum of shallow-water waves on the sphere is purely real. We now introduce the Weyl symbol of the wave operator $\mathcal{H}$:
\begin{equation} \label{eq:symbol}
    H = \begin{pmatrix} 0 && \ii f && M \\[6pt] -\ii f && 0 && K_\theta - \ii \beta_{\rm g} \\[6pt] M && K_\theta + \ii \beta_{\rm g} && 0 \end{pmatrix} \ ,
\end{equation}
where we have also defined $M = c \: m / (R \cos \theta)$, and $K_\theta$ the symbol of the Hermitian differential operator $-\ii \dd / \dd \theta$ times $c/R$, which is therefore a real quantity (see \cite{hall2013quantum}, Theorem 13.8). Generally speaking, the symbol of a wave operator, which is obtained via the Wigner transform (see appendix \ref{apx:WW}), provides a local representation of the wave equations in phase space (see e.g. \cite{littlejohn1991geometric,onuki2020quasi,perez2021manifestation,venaille2023ray}). It extends the Fourier transform to systems of wave equations with non constant coefficients. In the present case, the symbol $H$ represents quasi-local dispersion and polarisation relations of shallow-water waves on the sphere. $H$ is a continuous matrix function of the conjugated variables $\theta, K_\theta$, and $m$. To be clear, $m$ must be an integer so that the wave functions of $\mathcal{H}$ are regular on the sphere, but we can formally consider the symbol $H$ as a continuous function of $m$ or $M$. In the context of this paper, we introduce it in order to apply the index theorem, following \cite{faure2019manifestation}, i.e. we investigate the topological properties of the degeneracy points of $H$. This theorem relates the Chern numbers of these degeneracy points to the number and direction of transiting modes of the wave operator $\mathcal{H}$, which can be interpreted as the spectral footprint of a degeneracy point of non-zero Chern number in the quasi-local dispersion relations. The symbol \eqref{eq:symbol} is a generalisation of the one studied by \cite{delplace2017} for the unbounded $\beta$-plane: indeed, in the Matsuno limit, $\beta_{\rm g} / f \approx \epsilon/4 \ll 1$ in the Yoshida waveguide, thus the metric term $\beta_{\rm g}$ can be dropped, as far as equatorially-trapped waves are concerned.\\

A recent paper \cite{ageev2024unveiling} wrongly states that the shallow-water wave operator on the sphere is non-Hermitian and admits complex eigenvalues, which is not true since this system is linearly stable in the absence of a background flow. The mistake originates from using the Fourier transform (which is wrong since the coefficients of $\mathcal{H}$ vary with $\theta$) on the wave operator, instead of the more relevant Wigner transform. It is known that the Wigner transform must include additional terms on curved manifolds \cite{gneiting2013quantum}, which yields a Hermitian symbol in this case. Moreover, the equivalence between the respective Hermiticity of an operator and its symbol is a known property \cite{hall2013quantum}, and the symbol \eqref{eq:symbol} is Hermitian as we used an appropriate rescaling of the variables, implying the Hermiticity of the operator $\mathcal{H}$ for the canonical scalar product. Besides, we wish to point out that the topological properties that are presented in the following section do not depend on the choice of $\theta$ (and thus its conjugated symbol $K_\theta$) as coordinate to express the differential operator $\mathcal{H}$ and its symbol. Indeed, one would obtain the same symbol $H$ using any alternative coordinate $x=F(\theta)$, as long as the fields are appropriately rescaled so as to preserve the Hermiticity of $\mathcal{H}$ (see appendix \ref{apx:WW}).\\

We also wish to point out that Equations \eqref{eq:Schrodinger_SW} are formally equivalent to the shallow-water model in planar geometry with varying topography, whose transiting modes were investigated by \cite{venaille2021}. The metric term $\beta_{\rm g}$ plays the exact same role as the topographic parameter $\beta_t$, although \cite{venaille2021} only considered an $f$-plane. In other words, the metric $\beta$-term is mathematically analogous to topography in flat metric. Actually, one could consider together the effect of a curved metric (spherical or oblate for rapidly rotating celestial bodies, for instance) with a topography $h_0 (\theta)$ varying with latitude (thus the velocity $c=\sqrt{gh_0}$ is also a function of latitude). It can be shown that the symbol of this problem is the same as \eqref{eq:symbol}, with a total $\beta$-term that is the sum of $\beta_{\rm g}$ and $\beta_t$. For the spherical metric with topography, this adds up to $\beta_\text{total} = (c/2R) \left( \tan \theta - \dd \ln c / \dd \theta \right)$. We will not be considering a varying topography, as we wish to focus on the curved metric. However, we bring to the analysis of \cite{venaille2021} the case of varying $f$ and the additional geometric/topographic term with latitude, which was already addressed by \cite{perez2022topological} (part 3.5.1), in the context of the equatorial channel.

\subsection{Degeneracy points of the symbol and their Chern numbers} \label{part:eigenbands}

Let us now determine the topological properties of the matrix $H$ of Expression \eqref{eq:symbol}. As a function of $(m,K_\theta,\theta)$ and a 3-by-3 Hermitian matrix, $H$ has generically 3 eigenvalues which are real-valued functions defined over the parameter space $(m,K_\theta,\theta)$. These will be referred to as \textit{eigenbands} in the following, noted $\lambda_n$ with $n=-1,0,+1$ for increasing eigenvalues. It is important to understand that these eigenvalues are not the same as the frequencies $\omega$, i.e. the eigenvalues of the operator $\mathcal{H}$ as defined in \eqref{eq:Schrodinger_SW}.\\

The fields in Equations \eqref{eq:linearised_SW} are real-valued. As such, we have $\overline{\mathcal{H}(m)} = -\mathcal{H}(-m)$, where the overline stands for the complex conjugation of all coefficients of $\mathcal{H}$. Consequently, the spectrum of $\mathcal{H}$ is completely symmetric under the inversion $(m , \omega) \rightarrow (-m , -\omega)$. In the same way, all eigenvalues of $H$ verify
\begin{equation}
    \lambda_{-n} (m,K_\theta,\theta) = -\lambda_{+n} (-m,-K_\theta,\theta) \ ,
\end{equation}
for any band index $n=-1,0,+1$. Therefore, it is sufficient to consider only the positive frequencies $\omega$ of $\mathcal{H}$, and the eigenbands $n=0$ (quasi-local representation of Rossby waves) and $n=+1$ (quasi-local representation of inertia-gravity waves) of $H$. For this reason, even though a given degeneracy point of $H$ has three Chern numbers (one for each eigenband $n=-1,0,+1$), we will mostly consider the $n=+1$ one and simply refer to it as Chern number (without mentioning the band index). The characteristic polynomial of the matrix $H$ reads as
\begin{equation} \label{eq:characteristic_polynomial}
    X^3 - \left( M^2 + K_{\theta}^2 + f^2 + \beta_{\rm g}^2 \right) X - 2 f \beta_{\rm g} M \ .
\end{equation}

The polynomial \eqref{eq:characteristic_polynomial} has a multiple root, i.e. $H$ has a degenerated eigenvalue ($\lambda_{0} = \lambda_{+1}$ and/or $\lambda_{0} = \lambda_{-1}$), if and only if
\begin{equation}
    \beta_{\rm g} = \pm f \ ,
\end{equation}
as computed by \cite{venaille2021} for the topographic shallow-water model. The degenerated positive eigenvalue ($\lambda_{0} = \lambda_{+1}$) is equal to $|f|$, and the corresponding degeneracy points in parameter space are the ones such that $\beta_{\rm g} (\theta) = \pm f (\theta)$, $M = \mp |f|$ and $K_\theta = 0$. Now the modal flow of $\mathcal{H}$ for an eigenband $n$, i.e. the number of modes gained by the Rossby (for $n=0$) or the inertia-gravity (for $n=+1$) waveband as $m$ goes from $-\infty$ to $+\infty$, is constrained by the presence of at least one of these degeneracy points in the real system, and the value of this modal flow is given by the Chern number (a negative Chern number corresponds to a net loss of modes as $m$ increases) held by this degeneracy point for the band $n$ of the symbol \cite{delplace2017,faure2019manifestation,perez2022topological}. Since $f$ and $\beta_{\rm g}$ have the same sign (negative in the southern hemisphere and positive in the northern one), the problem amounts to solving $c \tan (\theta) / 2 R = 2 \Omega \sin (\theta)$, which yields
\begin{equation}
    \theta = 0 \quad \text{or} \quad \cos (\theta) = \frac{\epsilon}{4} \ .
\end{equation}

\begin{figure}[H]
    \centering
    \includegraphics[scale=0.53]{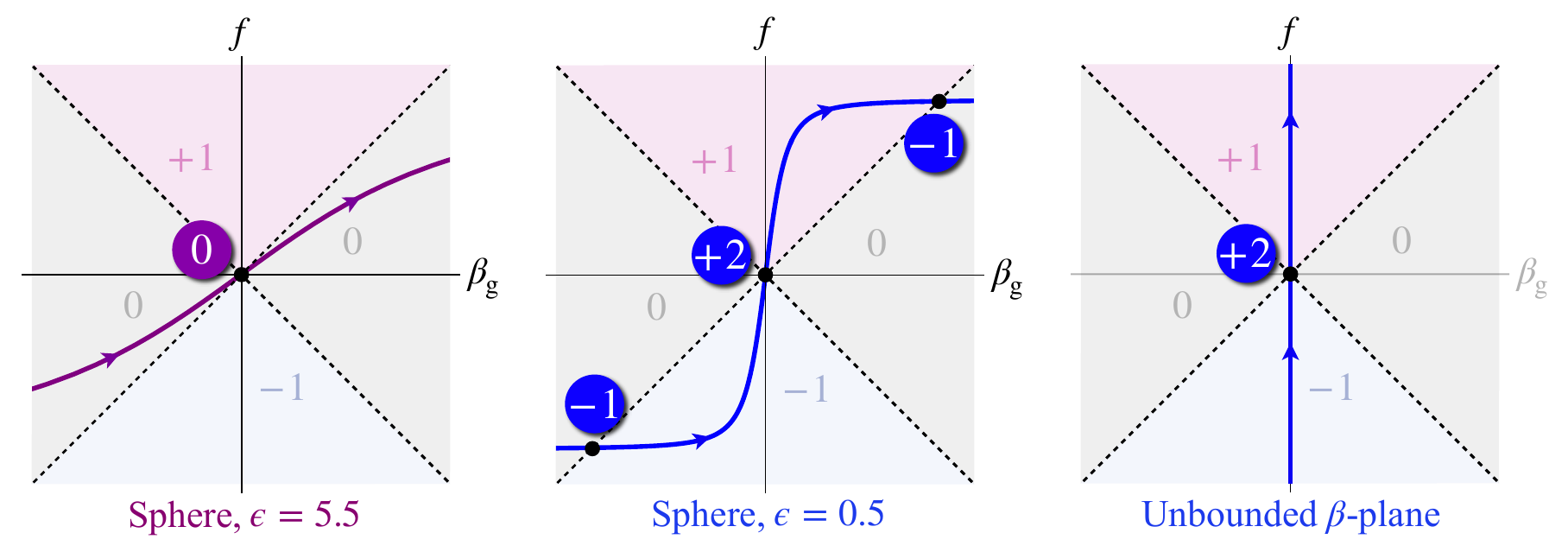}
    \caption{\label{fig:phase_diagram} Phase diagram of $H$, inspired by \cite{haldane1988model}. The colored curves represent the trajectory of $(\beta_{\rm g} , f)$ as $\theta$ goes from $-\pi/2$ to $+\pi/2$. $H$ has a degenerated eigenvalue when $f = \pm \beta_{\rm g}$ (dashed lines), thus splitting the parameter space $(\beta_{\rm g} , f)$ in 4 areas, each of which is characterised by an integer $0,-1$ or $+1$. The Chern number $C_{+1}$ (colored discs) of a degeneracy point (black dots) involving the eigenband $n=+1$ is given by their difference as $\theta$ increases past the degeneracy point. Middle: if $\epsilon < 4$, the equatorial degeneracy point has $C_{+1}=+2$, same as in \cite{delplace2017}, and there are two additional mid-latitude degeneracy points with $C_{+1}=-1$. Left: as $\epsilon$ increases and reaches $4$, they merge with the equatorial degeneracy point, which still exists for $\epsilon > 4$ and has zero Chern number. Right: the symbol of shallow-water waves on the unbounded $\beta$-plane is obtained by taking $\beta_{\rm g} = 0$ and $f=\beta y$ in \eqref{eq:symbol}, which yields the unique degeneracy point of Chern number $+2$ studied by \cite{delplace2017}.}
\end{figure}

Therefore one can distinguish two situations, as illustrated in Figure \ref{fig:phase_diagram}:

\begin{itemize}
    \item If $\epsilon < 4$, there are three latitudes at which a degeneracy between the eigenvalues of the symbol occurs, one at the equator (3-fold since all three eigenvalues of $H$ are degenerated there) and two (2-fold) degeneracy points near the poles, at opposite latitude. As $\epsilon$ increases, these get closer to the equator, where they eventually merge with the equatorial degeneracy point for $\epsilon = 4$. The value of $m$ at the non-equatorial degeneracy points for the positive eigenbands is
    \begin{equation} \label{eq:m_c}
        m_c = -\frac{1}{2} \sqrt{1 - \left( \frac{\epsilon}{4} \right)^2} \: \in (-1/2 , 0) \ .
    \end{equation}
    
    \item If $\epsilon \geq 4$, there is a unique degeneracy point at the equator. This can be seen as a threshold in the competition between two different $\beta$-effects, the traditional one owing to the variation of $f$ with latitude, which tends to trap waves at the equator, and the geometric one, which conversely tends to spread the wave functions away from it, over the whole sphere.
\end{itemize}

Each of these degeneracy points can be assigned a set of Chern numbers, one for each eigenband involved in the degeneracy (see Appendix \ref{apx:topology}). Again, two situations can be distinguished:

\begin{itemize}
    \item For $\epsilon < 4$, the equatorial 3-fold degeneracy point has a set of Chern numbers $(C_{-1},C_{0},C_{+1}) = (-2,0,+2)$ for the three eigenbands ($n=-1,0,+1$) of $H$, which are all degenerated. This is the result of \cite{delplace2017} for the $\beta$-plane shallow-water model, and it still holds with the contribution of the spherical metric, as long as $\epsilon < 4$. The other 2-fold non-equatorial degeneracy points bear the Chern numbers $(C_{0},C_{+1}) =(+1,-1)$ (for the positive-eigenband degeneracies) and $(C_{-1},C_{0}) =(+1,-1)$ (for the negative-eigenband degeneracies). Note that, for small $\epsilon$, the latter are pushed toward the poles, which are then discarded by Matsuno's $\beta$-plane approximation. This is why the $\beta$-plane shallow-water model does not have them accounted for. These negatively-charged degeneracy points also appear at wave number $m = m_c \in (-1/2 , 0)$, given by Expression \eqref{eq:m_c}, which is why their footprint is appreciable in the form of a jump of the modal branches right before $m$ reaches $0$ (see Figures \ref{fig:branches_v_label} and \ref{fig:modes_0.2}).
    
    \item For $\epsilon \geq 4$, all degeneracy points merge at the equator -- and their respective Chern numbers add up -- into a unique three-fold degeneracy point which is topologically neutral, i.e. of zero Chern numbers: $(C_{-1},C_{0},C_{+1}) = (0,0,0)$. In the spectrum of the wave operator, this fusion manifests as a symmetrisation of the modes between positive and negative $m$, whereas the east-west asymmetry of the spectrum is appreciable for small $\epsilon$, when the degeneracy points are separated.
\end{itemize}

In conclusion, the metric term $\beta_{\rm g}$ generates systematic degeneracy points whose Chern numbers compensate that of the equatorial degeneracy point. For small $\epsilon$, these degeneracy points are located near the poles and bear the Chern number $-1$ for the eigenband $n=+1$, and for $\epsilon \geq 4$ there is a unique degeneracy point at the equator whose Chern numbers are $0$. The eigenbands of $H$ thus have zero total Chern numbers, which confirms that the net number of modes gained by the inertia-gravity waveband as $m$ is swept from $-\infty$ to $+\infty$ is zero, in virtue of the index theorem. As explained in section \ref{sec:zeros}, there is a direct correspondence between this mode imbalance and the evolution of the number of zeros of $\Tilde{v}$ for waves on the unbounded $\beta$-plane. While the same correspondence can only be inferred by extrapolation for waves on the rotating sphere, we showed here that the absence of modal flow is in agreement with the predictions of topology. For the unbounded $\beta$-plane, \cite{venaille2023ray} provides a formal proof of the equality between the modal flow (i.e. the difference in number of positive-frequency inertia-gravity modes between the two limits $k \rightarrow + \infty$ and $k \rightarrow -\infty$) and the Chern number of the symbol's degeneracy point at the equator, based on geometric WKB analysis of the waves in both limits. This study, which provides an alternative point of view to the index theorem, could also be extended here to demonstrate the nullity of this mode imbalance on the sphere, with the wave number $m$ instead of $k$.

\subsection{Collapse of the spectral flow} \label{part:collapse}

For $\epsilon < 4$, one could expect that each of the degeneracy points exhibited in \ref{part:eigenbands} should be associated with its own branch of modes, two of them localised near the equator, transiting from the Rossby waveband to the inertia-gravity waveband as $m$ increases (corresponding to the $+2$ topological charge at the equator), and two localised near the latitudes $\pm \theta$ such that $\cos \theta = \epsilon / 4$, transiting from the inertia-gravity waveband to the Rossby waveband as $m$ increases (corresponding to the $-1$ charges). However there is no footprint of the modes transiting from the inertia-gravity waveband to the Rossby waveband, and this happens for two reasons:

\begin{itemize}
    \item Although the index theorem applies individually for each degeneracy point of the symbol, the resulting spectral flows are visible only in a semi-classical limit \cite{faure2019manifestation}, i.e. as long as the transiting modes are well-separated in phase space \cite{jezequel23}. In the case of shallow-water waves on the rotating sphere, the semi-classical limit in question corresponds to $\epsilon \ll 1$. Indeed, the equatorial waves spread up to the poles even for moderate values of $\epsilon$ ($\epsilon \sim 1$), which means that the modes corresponding to the opposite spectral flows overlap and hybridise, thus manifesting an avoided crossing between their respective spectral branches (see e.g. the cases studied by \cite{kaufman1999mode} and \cite{perez2022topological}, part 3.5.1), which is precisely why the spectral flow of frequencies collapses as $\epsilon$ increases.
    
    \item Even for $\epsilon \ll 1$ the modes of negative spectral flow are not visible in the spectrum. We showed in \ref{part:eigenbands} that these are expected to be localised near the poles, as the corresponding degeneracy points of the symbol are, and transit around $m \approx -1/2$. These modes are the spherical counterpart of Matsuno's spurious solutions (for $\omega = -c k$), or coastal Kelvin waves in the equatorial channel \cite{paldor2015,venaille2021,perez2022topological}, if the edges of the channel were moved up to the poles. Owing to the spherical geometry, the dispersion relation of coastal Kelvin modes propagating along a longitudinal wall at latitude $\theta$ is $\omega = -cm/R \cos \theta$. Therefore, the slope of the branches formed by these hypothetical polar modes is large, and they do not appear in the spectrum for integer values of the wave number $m$ since they transit around $m \approx -1/2$. For small $\epsilon$, the strong east-west asymmetry is the manifestation of these opposite topological charges being separated, the gap-crossing Kelvin and Yanai frequencies are the footprint of the $+2$ topological charge at the equator, while the abrupt $+2$ modal jump happening between $m=-1$ and $m=0$ (see Figure \ref{fig:branches_v_label}) is the footprint of the degeneracy points of charges $-1$ located near the poles.
\end{itemize}

\section{Concluding remarks}

The goal of this work is to investigate the topological properties of the shallow-water spectrum on the rotating sphere. In particular, we aimed at answering the following questions: how do the results of \cite{delplace2017}, which were established using the $\beta$-plane approximation, extend to the spherical case? What does topology tell us about the transition between the Matsuno and Margules limits? Both regimes have been extensively investigated, and the corresponding solutions can be well-approximated, however there is no analytic solution for arbitrary values of $\epsilon$. In the Matsuno limit ($\epsilon \ll 1$ or large Lamb parameter), there seems to be a continuous spectral connection between the Rossby and inertia-gravity modes, which is embodied by the Yanai and Kelvin modes, whose frequencies cross the gap as $m$ increases (see Figure \ref{fig:asymptotic_regimes}). This spectral feature is reminiscent of the Matsuno spectrum on the unbounded $\beta$-plane, for which the connection with topological invariant is well-established \cite{delplace2017,faure2019manifestation,delplace2022berry,venaille2023ray}. In the Margules limit (large $\epsilon$ or small Lamb parameter), this gapless connection progressively breaks down. All the inertia-gravity modes plus the positive-phase-speed Yanai (with $m \geq 0$) and Kelvin modes (with $m \geq 1$) move up to values $\omega = \mathcal{O}(c/R)$, while the largest Yanai frequency ($m=-1$) remains smaller than $\Omega$ (see Equation \eqref{eq:Rossby_sphere} with $n=0$ and Figure \ref{fig:transition}), thus separating the spectrum into two distinct wavebands with an open frequency gap.\\

For small $\epsilon$, the frequencies of shallow-water waves on the sphere and those of the Matsuno spectrum on the unbounded $\beta$-plane are nearly identical, however the $\Tilde{v}$ component of the modes of positive (eastward) phase speed has two more zeros located at opposite latitudes. The latter being close to the poles, they are not captured by the $\beta$-plane approximation. Modes on the sphere have been extensively studied, in particular the Yanai modes \cite{garfinkel2017classification,paldor2018mixed}. Nevertheless, to our knowledge, the existence of these additional zeros is usually not discussed in the geophysical literature. Yet they are consistent with the evolution of the shallow-water spectrum as $\epsilon$ varies continuously. To explain why, let us consider a mode with wave number $m$ and frequency $\omega$, for a large value of $\epsilon$ (Margules limit). The number $p$ of zeros of $\Tilde{v}(\theta)$ is established by the classification given in \ref{part:asymptotic_solutions}. The inertia-gravity modes are essentially those of the Laplacian on the sphere, which are insensitive to the sign of $m$. As $\epsilon$ continuously decreases, the effect of rotation becomes stronger and the spectrum progressively loses this east-west symmetry. For the mode picked in the large $\epsilon$ regime, its frequency $\omega$ and the wave function $\Tilde{v}$ change as well in a continuous manner. However, the number $p$ cannot change as the structure of $\Tilde{v}$ is governed by a generalised Sturm-Liouville problem (see the arguments of \cite{iga1995transition}), whose eigenvalue equation is the spherical generalisation of Matsuno's equation (Equation (3.3) of \cite{muller1995shallow} is given for the perturbation of potential vorticity, which is proportional to $\cos{(\theta)} \Tilde{v}$). When decreasing $\epsilon$, the frequencies of the eastward modes decrease faster than the westward ones, leading to a dislocation (or $+2$ branch jump) in the spectrum. This dislocation eventually results in the spectral flow of Yanai and Kelvin modes in the small $\epsilon$ (Matsuno) limit. In the Matsuno limit on the sphere, this spectral flow is thus obtained as a continuous deformation of the Margules spectrum with no spectral flow. It arises from the isolated $+2$ topological charge at the equator, whereas the opposite topological charges near the poles generate invisible modes with opposite spectral flow, which manifest in the spectrum through the change in number of $\Tilde{v}$ zeros of Yanai modes as $m$ changes sign. In other words, this spectral flow is not a modal flow, since the westward and eastward Yanai modes do not constitute a single modal branch. In contrast, the shallow-water spectrum on the unbounded $\beta$-plane arises from a single non-zero topological charge, because there is no such way to continuously deform it into a spectrum with no modal flow, as the parameters $\beta$ and $c$ can be absorbed in the definition of length and time units, and thus do not affect the shape of the spectrum discussed in \ref{part:Matsuno}. Conversely, for waves on the sphere, both $R$ and the Rossby radius of deformation $c/2\Omega$ deeply affect the shape of the solutions, and their ratio define the effective equatorial trapping of waves.\\

We then highlighted furthermore this distinction between the unbounded $\beta$-plane and the sphere by investigating the topology of the symbol of the shallow-water model on the sphere. We showed that the shallow-water eigenbands on the sphere have zero total Chern numbers, in contrast with the eigenbands on the unbounded $\beta$-plane. Consequently, in virtue of the index theorem, there is no net modal flow of modes in the spectrum. Precisely, we found that when $\epsilon$ becomes higher than $4$, several degeneracy points merge into a unique, topologically neutral degeneracy point at the equator. This manifests in the shallow-water spectrum as the loss of east-west asymmetry and the complete collapse of the spectral flow of equatorial Yanai and Kelvin waves. However, for $\epsilon < 4$, the symbol of the wave operator on the sphere has several charged degeneracy points: an equatorial one with non-zero Chern numbers, which is reminiscent of the unbounded $\beta$-plane \cite{delplace2017}, plus two others (considering the positive eigenbands $n=0,+1$) with opposite charges. As discussed in \ref{part:collapse}, the spectral footprint of the former are the Yanai and Kelvin waves, whose transiting frequencies are characteristic of the east-west asymmetry at small $\epsilon$, and that of the latter is the $+2$ modal jump, i.e. the fact that opposite topological charges of the symbol lead, for small $\epsilon$, to a spectral flow that is not a modal flow.\\

This work joins recent efforts toward the comprehension of how topology applies in the spectral properties of continuous physical systems on curved surfaces \cite{shankar2017topological,green2020topological,finnigan2022equatorial,li2023,ageev2024unveiling}, a subject that still lacks a unified framework. In that sense, we wish to stress the importance of the Weyl symbol and the spectral parameter. With this study, we have shown that the shallow-water model on the rotating sphere has a counter-intuitive topology that is either neutral (all Chern numbers are null for $\epsilon \geq 4$) or \textit{polar} in the sense that it has several degeneracy points of non-zero Chern numbers which sum to zero (for $\epsilon < 4$), whereas its $\beta$-plane counterpart in unbounded flat geometry has a unique, topologically charged degeneracy point. Besides, we exhibited a peculiar situation, in which different degeneracy points (with different multiplicity) merge and yield a unique degeneracy point which is topologically neutral, yet without the gap opening afterward. This situation is not usual in topological physics. Further investigation is necessary to better understand the spectral properties of continuous systems on curved surfaces, and the manifestation of topology in their dynamics.\\

NP and GL acknowledge funding from the ERC CoG project PODCAST No 864965. AL is funded by a PhD grant allocation Contrat doctoral Normalien. PD is supported by the national grant ANR-18-CE30-0002-01.

\bibliography{biblio}

\begin{thebibliography}{10}

\bibitem{delplace2017}
P.~Delplace, J.~Marston, and A.~Venaille, ``Topological origin of equatorial waves,'' {\em Science}, vol.~358, no.~6366, pp.~1075--1077, 2017.

\bibitem{gill1982atmosphere}
A.~E. Gill, {\em Atmosphere-ocean dynamics}, vol.~30.
\newblock Academic press, 1982.

\bibitem{vallis2017atmospheric}
G.~K. Vallis, {\em Atmospheric and oceanic fluid dynamics}.
\newblock Cambridge University Press, 2017.

\bibitem{zeitlin2018geophysical}
V.~Zeitlin, {\em GEOPHYSICAL FLUID DYNAMICS: Understanding (almost) everything with rotating shallow water models}.
\newblock Oxford University Press, 2018.

\bibitem{gilman2000magnetohydrodynamic}
P.~A. Gilman, ``Magnetohydrodynamic “shallow water” equations for the solar tachocline,'' {\em The Astrophysical Journal}, vol.~544, no.~1, p.~L79, 2000.

\bibitem{zaqarashvili2009}
T.~Zaqarashvili, R.~Oliver, and J.~Ballester, ``Global shallow water magnetohydrodynamic waves in the solar tachocline,'' {\em The Astrophysical Journal}, vol.~691, no.~1, p.~L41, 2009.

\bibitem{zaqarashvili2021rossby}
T.~Zaqarashvili, M.~Albekioni, J.~Ballester, Y.~Bekki, L.~Biancofiore, A.~Birch, M.~Dikpati, L.~Gizon, E.~Gurgenashvili, E.~Heifetz, {\em et~al.}, ``Rossby waves in astrophysics,'' {\em Space Science Reviews}, vol.~217, pp.~1--93, 2021.

\bibitem{thomson18801}
W.~Thomson, ``1. on gravitational oscillations of rotating water,'' {\em Proceedings of the Royal Society of Edinburgh}, vol.~10, pp.~92--100, 1880.

\bibitem{rossby1939relation}
C.-G. Rossby, ``Relation between variations in the intensity of the zonal circulation of the atmosphere and the displacements of the semi-permanent centers of action,'' {\em J. mar. Res.}, vol.~2, pp.~38--55, 1939.

\bibitem{stommel1948westward}
H.~Stommel, ``The westward intensification of wind-driven ocean currents,'' {\em Eos, Transactions American Geophysical Union}, vol.~29, no.~2, pp.~202--206, 1948.

\bibitem{munk1950wind}
W.~H. Munk and G.~F. Carrier, ``The wind-driven circulation in ocean basins of various shapes,'' {\em Tellus}, vol.~2, no.~3, pp.~158--167, 1950.

\bibitem{rossby1948displacements}
C.~Rossby, ``On displacements and intensity changes of atmospheric vortices,'' {\em Journal of Marine Research}, vol.~7, no.~3, 1948.

\bibitem{matsuno1966}
T.~Matsuno, ``Quasi-geostrophic motions in the equatorial area,'' {\em Journal of the Meteorological Society of Japan. Ser. II}, vol.~44, no.~1, pp.~25--43, 1966.

\bibitem{yoshida1959theory}
K.~Yoshida, ``A theory of the cromwell current (the equatorial undercurrent) and of the equatorial upwelling an interpretation in a similarity to a costal circulation,'' {\em Journal of the Oceanographical Society of Japan}, vol.~15, no.~4, pp.~159--170, 1960.

\bibitem{margules1980air}
M.~Margules, {\em Air motions in a rotating spheroidal shell}.
\newblock Advanced Study Program, National Center for Atmospheric Research, 1980.

\bibitem{hough1898v}
S.~S. Hough, ``V. on the application of harmonic analysis to the dynamical theory of the tides.—part ii. on the general integration of laplace’s dynamical equations,'' {\em Philosophical Transactions of the Royal Society of London. Series A, Containing Papers of a Mathematical or Physical Character}, vol.~191, pp.~139--185, 1898.

\bibitem{longuet1968}
M.~S. Longuet-Higgins, ``The eigenfunctions of laplace's tidal equation over a sphere,'' {\em Philosophical Transactions of the Royal Society of London. Series A, Mathematical and Physical Sciences}, vol.~262, no.~1132, pp.~511--607, 1968.

\bibitem{bridger1980long}
A.~F. Bridger and D.~E. Stevens, ``Long atmospheric waves and the polar-plane approximation to the earth’s spherical geometry,'' {\em Journal of the Atmospheric Sciences}, vol.~37, no.~3, pp.~534--544, 1980.

\bibitem{muller1995shallow}
D.~M{\"u}ller and J.~O’Brien, ``Shallow water waves on the rotating sphere,'' {\em Physical Review E}, vol.~51, no.~5, p.~4418, 1995.

\bibitem{dellar2011variations}
P.~J. Dellar, ``Variations on a beta-plane: derivation of non-traditional beta-plane equations from hamilton's principle on a sphere,'' {\em Journal of Fluid Mechanics}, vol.~674, pp.~174--195, 2011.

\bibitem{paldor2015}
N.~Paldor, {\em Shallow water waves on the rotating Earth}.
\newblock Springer, 2015.

\bibitem{vasil2019tensor}
G.~M. Vasil, D.~Lecoanet, K.~J. Burns, J.~S. Oishi, and B.~P. Brown, ``Tensor calculus in spherical coordinates using jacobi polynomials. part-i: Mathematical analysis and derivations,'' {\em Journal of Computational Physics: X}, vol.~3, p.~100013, 2019.

\bibitem{faure2019manifestation}
F.~Faure, ``Manifestation of the topological index formula in quantum waves and geophysical waves,'' {\em Annales Henri Lebesgue}, vol.~6, pp.~449--492, 2023.

\bibitem{delplace2022berry}
P.~Delplace, ``Berry-chern monopoles and spectral flows,'' {\em SciPost Physics Lecture Notes}, p.~039, 2022.

\bibitem{QinFu22}
H.~Qin and Y.~Fu, ``Topological langmuir-cyclotron wave,'' {\em Science Advances}, vol.~9, no.~13, p.~eadd8041, 2023.

\bibitem{perrot2019topological}
M.~Perrot, P.~Delplace, and A.~Venaille, ``Topological transition in stratified fluids,'' {\em Nature Physics}, vol.~15, no.~8, pp.~781--784, 2019.

\bibitem{venaille2021}
A.~Venaille and P.~Delplace, ``Wave topology brought to the coast,'' {\em Physical Review Research}, vol.~3, no.~4, p.~043002, 2021.

\bibitem{perez2022unidirectional}
N.~Perez, P.~Delplace, and A.~Venaille, ``Unidirectional modes induced by nontraditional coriolis force in stratified fluids,'' {\em Physical Review Letters}, vol.~128, no.~18, p.~184501, 2022.

\bibitem{leclerc2022topological}
A.~Leclerc, G.~Laibe, P.~Delplace, A.~Venaille, and N.~Perez, ``Topological modes in stellar oscillations,'' {\em The Astrophysical Journal}, vol.~940, no.~1, p.~84, 2022.

\bibitem{parker2020topological}
J.~B. Parker, J.~Marston, S.~M. Tobias, and Z.~Zhu, ``Topological gaseous plasmon polariton in realistic plasma,'' {\em Physical Review Letters}, vol.~124, no.~19, p.~195001, 2020.

\bibitem{qin2023topological}
H.~Qin and Y.~Fu, ``Topological langmuir-cyclotron wave,'' {\em Science Advances}, vol.~9, no.~13, p.~eadd8041, 2023.

\bibitem{wang2009observation}
Z.~Wang, Y.~Chong, J.~D. Joannopoulos, and M.~Solja{\v{c}}i{\'c}, ``Observation of unidirectional backscattering-immune topological electromagnetic states,'' {\em Nature}, vol.~461, no.~7265, pp.~772--775, 2009.

\bibitem{souslov2017topological}
A.~Souslov, B.~C. Van~Zuiden, D.~Bartolo, and V.~Vitelli, ``Topological sound in active-liquid metamaterials,'' {\em Nature Physics}, vol.~13, no.~11, pp.~1091--1094, 2017.

\bibitem{shankar2017topological}
S.~Shankar, M.~J. Bowick, and M.~C. Marchetti, ``Topological sound and flocking on curved surfaces,'' {\em Physical Review X}, vol.~7, no.~3, p.~031039, 2017.

\bibitem{nash2015topological}
L.~M. Nash, D.~Kleckner, A.~Read, V.~Vitelli, A.~M. Turner, and W.~T. Irvine, ``Topological mechanics of gyroscopic metamaterials,'' {\em Proceedings of the National Academy of Sciences}, vol.~112, no.~47, pp.~14495--14500, 2015.

\bibitem{khanikaev2015topologically}
A.~B. Khanikaev, R.~Fleury, S.~H. Mousavi, and A.~Alu, ``Topologically robust sound propagation in an angular-momentum-biased graphene-like resonator lattice,'' {\em Nature communications}, vol.~6, no.~1, p.~8260, 2015.

\bibitem{perez2022topological}
N.~Perez, {\em Topological waves in geophysical and astrophysical fluids}.
\newblock PhD thesis, Ecole normale sup{\'e}rieure de lyon-ENS LYON, 2022.

\bibitem{aerts2010asteroseismology}
C.~Aerts, J.~Christensen-Dalsgaard, and D.~W. Kurtz, {\em Asteroseismology}.
\newblock Springer Science \& Business Media, 2010.

\bibitem{longuet1965planetary}
M.~S. Longuet-Higgins, ``Planetary waves on a rotating sphere. ii,'' {\em Proceedings of the Royal Society of London. Series A. Mathematical and Physical Sciences}, vol.~284, no.~1396, pp.~40--68, 1965.

\bibitem{muller1994}
D.~M{\"u}ller, B.~Kelly, and J.~O'brien, ``Spheroidal eigenfunctions of the tidal equation,'' {\em Physical review letters}, vol.~73, no.~11, p.~1557, 1994.

\bibitem{tacseli2003exact}
H.~Ta{\c{s}}eli, ``Exact analytical solutions of the hamiltonian with a squared tangent potential,'' {\em Journal of mathematical chemistry}, vol.~34, pp.~243--251, 2003.

\bibitem{paldor2018mixed}
N.~Paldor, I.~Fouxon, O.~Shamir, and C.~I. Garfinkel, ``The mixed rossby--gravity wave on the spherical earth,'' {\em Quarterly Journal of the Royal Meteorological Society}, vol.~144, no.~715, pp.~1820--1830, 2018.

\bibitem{burns2020}
K.~J. {Burns}, G.~M. {Vasil}, J.~S. {Oishi}, D.~{Lecoanet}, and B.~P. {Brown}, ``{Dedalus: A flexible framework for numerical simulations with spectral methods},'' {\em Physical Review Research}, vol.~2, p.~023068, Apr. 2020.

\bibitem{tan2020}
X.~Tan and A.~P. Showman, ``Atmospheric circulation of tidally locked gas giants with increasing rotation and implications for white dwarf--brown dwarf systems,'' {\em The Astrophysical Journal}, vol.~902, no.~1, p.~27, 2020.

\bibitem{monnier2010}
J.~Monnier, R.~Townsend, X.~Che, M.~Zhao, T.~Kallinger, J.~Matthews, and A.~Moffat, ``Rotationally modulated g-modes in the rapidly rotating $\delta$ scuti star rasalhague ($\alpha$ ophiuchi),'' {\em The Astrophysical Journal}, vol.~725, no.~1, p.~1192, 2010.

\bibitem{showman2010atmospheric}
A.~P. Showman, J.~Y. Cho, and K.~Menou, ``Atmospheric circulation of exoplanets,'' {\em Exoplanets}, vol.~526, pp.~471--516, 2010.

\bibitem{shamir2023matsuno}
O.~Shamir, C.~I. Garfinkel, E.~P. Gerber, and N.~Paldor, ``The matsuno--gill model on the sphere,'' {\em Journal of Fluid Mechanics}, vol.~964, p.~A32, 2023.

\bibitem{del1990}
A.~D. Del~Genio and W.~B. Rossow, ``Planetary-scale waves and the cyclic nature of cloud top dynamics on venus,'' {\em Journal of Atmospheric Sciences}, vol.~47, no.~3, pp.~293--318, 1990.

\bibitem{johnson1993structure}
E.~S. Johnson and M.~J. Mc~Phaden, ``Structure of intraseasonal kelvin waves in the equatorial pacific ocean,'' {\em Journal of physical oceanography}, vol.~23, no.~4, pp.~608--625, 1993.

\bibitem{sprintall2000semiannual}
J.~Sprintall, A.~L. Gordon, R.~Murtugudde, and R.~D. Susanto, ``A semiannual indian ocean forced kelvin wave observed in the indonesian seas in may 1997,'' {\em Journal of Geophysical Research: Oceans}, vol.~105, no.~C7, pp.~17217--17230, 2000.

\bibitem{kiladis2009}
G.~N. Kiladis, M.~C. Wheeler, P.~T. Haertel, K.~H. Straub, and P.~E. Roundy, ``Convectively coupled equatorial waves,'' {\em Reviews of Geophysics}, vol.~47, no.~2, 2009.

\bibitem{sakazaki2020array}
T.~Sakazaki and K.~Hamilton, ``An array of ringing global free modes discovered in tropical surface pressure data,'' {\em Journal of the Atmospheric Sciences}, vol.~77, no.~7, pp.~2519--2539, 2020.

\bibitem{menou2009atmospheric}
K.~Menou and E.~Rauscher, ``Atmospheric circulation of hot jupiters: a shallow three-dimensional model,'' {\em The Astrophysical Journal}, vol.~700, no.~1, p.~887, 2009.

\bibitem{showman2018global}
A.~P. Showman, A.~P. Ingersoll, R.~Achterberg, and Y.~Kaspi, ``The global atmospheric circulation of saturn,'' {\em Saturn in the 21st Century}, vol.~20, p.~295, 2018.

\bibitem{gavriel2021number}
N.~Gavriel and Y.~Kaspi, ``The number and location of jupiter’s circumpolar cyclones explained by vorticity dynamics,'' {\em Nature geoscience}, vol.~14, no.~8, pp.~559--563, 2021.

\bibitem{legarreta2016}
J.~Legarreta, N.~Barrado-Izagirre, E.~Garc{\'\i}a-Melendo, A.~Sanchez-Lavega, and J.~M. G{\'o}mez-Forrellad, ``A large active wave trapped in jupiter’s equator,'' {\em Astronomy \& Astrophysics}, vol.~586, p.~A154, 2016.

\bibitem{zhu2023topology}
Z.~Zhu, C.~Li, and J.~Marston, ``Topology of rotating stratified fluids with and without background shear flow,'' {\em Physical Review Research}, vol.~5, no.~3, p.~033191, 2023.

\bibitem{iga1995transition}
K.~Iga, ``Transition modes of rotating shallow water waves in a channel,'' {\em Journal of Fluid Mechanics}, vol.~294, pp.~367--390, 1995.

\bibitem{tauber2019bulk}
C.~Tauber, P.~Delplace, and A.~Venaille, ``A bulk-interface correspondence for equatorial waves,'' {\em Journal of Fluid Mechanics}, vol.~868, p.~R2, 2019.

\bibitem{faure2000topological}
F.~Faure and B.~Zhilinskii, ``Topological chern indices in molecular spectra,'' {\em Physical review letters}, vol.~85, no.~5, p.~960, 2000.

\bibitem{fu2021topological}
Y.~Fu and H.~Qin, ``Topological phases and bulk-edge correspondence of magnetized cold plasmas,'' {\em Nature Communications}, vol.~12, no.~1, p.~3924, 2021.

\bibitem{venaille2023ray}
A.~Venaille, Y.~Onuki, N.~Perez, and A.~Leclerc, ``From ray tracing to waves of topological origin in continuous media,'' {\em SciPost Physics}, vol.~14, no.~4, p.~062, 2023.

\bibitem{hall2013quantum}
B.~C. Hall, {\em Quantum theory for mathematicians}.
\newblock Springer, 2013.

\bibitem{littlejohn1991geometric}
R.~G. Littlejohn and W.~G. Flynn, ``Geometric phases in the asymptotic theory of coupled wave equations,'' {\em Physical Review A}, vol.~44, no.~8, p.~5239, 1991.

\bibitem{onuki2020quasi}
Y.~Onuki, ``Quasi-local method of wave decomposition in a slowly varying medium,'' {\em Journal of Fluid Mechanics}, vol.~883, p.~A56, 2020.

\bibitem{perez2021manifestation}
N.~Perez, P.~Delplace, and A.~Venaille, ``Manifestation of the berry curvature in geophysical ray tracing,'' {\em Proceedings of the Royal Society A}, vol.~477, no.~2248, p.~20200844, 2021.

\bibitem{ageev2024unveiling}
D.~S. Ageev and A.~A. Iliasov, ``Unveiling topological modes on curved surfaces,'' {\em Physical Review B}, vol.~109, no.~8, p.~085435, 2024.

\bibitem{gneiting2013quantum}
C.~Gneiting, T.~Fischer, and K.~Hornberger, ``Quantum phase-space representation for curved configuration spaces,'' {\em Physical Review A}, vol.~88, no.~6, p.~062117, 2013.

\bibitem{haldane1988model}
F.~D.~M. Haldane, ``Model for a quantum hall effect without landau levels: Condensed-matter realization of the" parity anomaly",'' {\em Physical review letters}, vol.~61, no.~18, p.~2015, 1988.

\bibitem{jezequel23}
L.~Jezequel and P.~Delplace, ``Mode-shell correspondence, a unifying phase space theory in topological physics -- part i: Chiral number of zero-modes,'' {\em arXiv preprint arXiv:2310.05656}, 2023.

\bibitem{kaufman1999mode}
A.~Kaufman, J.~Morehead, A.~Brizard, and E.~Tracy, ``Mode conversion in the gulf of guinea,'' {\em Journal of Fluid Mechanics}, vol.~394, pp.~175--192, 1999.

\bibitem{garfinkel2017classification}
C.~I. Garfinkel, I.~Fouxon, O.~Shamir, and N.~Paldor, ``Classification of eastward propagating waves on the spherical earth,'' {\em Quarterly Journal of the Royal Meteorological Society}, vol.~143, no.~704, pp.~1554--1564, 2017.

\bibitem{green2020topological}
R.~Green, J.~Armas, J.~de~Boer, and L.~Giomi, ``Topological waves in passive and active fluids on curved surfaces: a unified picture,'' {\em arXiv preprint arXiv:2011.12271}, 2020.

\bibitem{finnigan2022equatorial}
C.~Finnigan, M.~Kargarian, and D.~K. Efimkin, ``Equatorial magnetoplasma waves,'' {\em Physical Review B}, vol.~105, no.~20, p.~205426, 2022.

\bibitem{li2023}
G.~Li and D.~K. Efimkin, ``Equatorial waves in rotating bubble-trapped superfluids,'' {\em Physical Review A}, vol.~107, no.~2, p.~023319, 2023.

\bibitem{berry1984quantal}
M.~V. Berry, ``Quantal phase factors accompanying adiabatic changes,'' {\em Proceedings of the Royal Society of London. A. Mathematical and Physical Sciences}, vol.~392, no.~1802, pp.~45--57, 1984.

\bibitem{fukui2005chern}
T.~Fukui, Y.~Hatsugai, and H.~Suzuki, ``Chern numbers in discretized brillouin zone: efficient method of computing (spin) hall conductances,'' {\em Journal of the Physical Society of Japan}, vol.~74, no.~6, pp.~1674--1677, 2005.

\end{thebibliography}

\section{Appendices}

\subsection{Numerically-computed frequencies and wave functions} \label{apx:numerics}

\begin{figure}[H]
    \centering
    \includegraphics[scale=0.5]{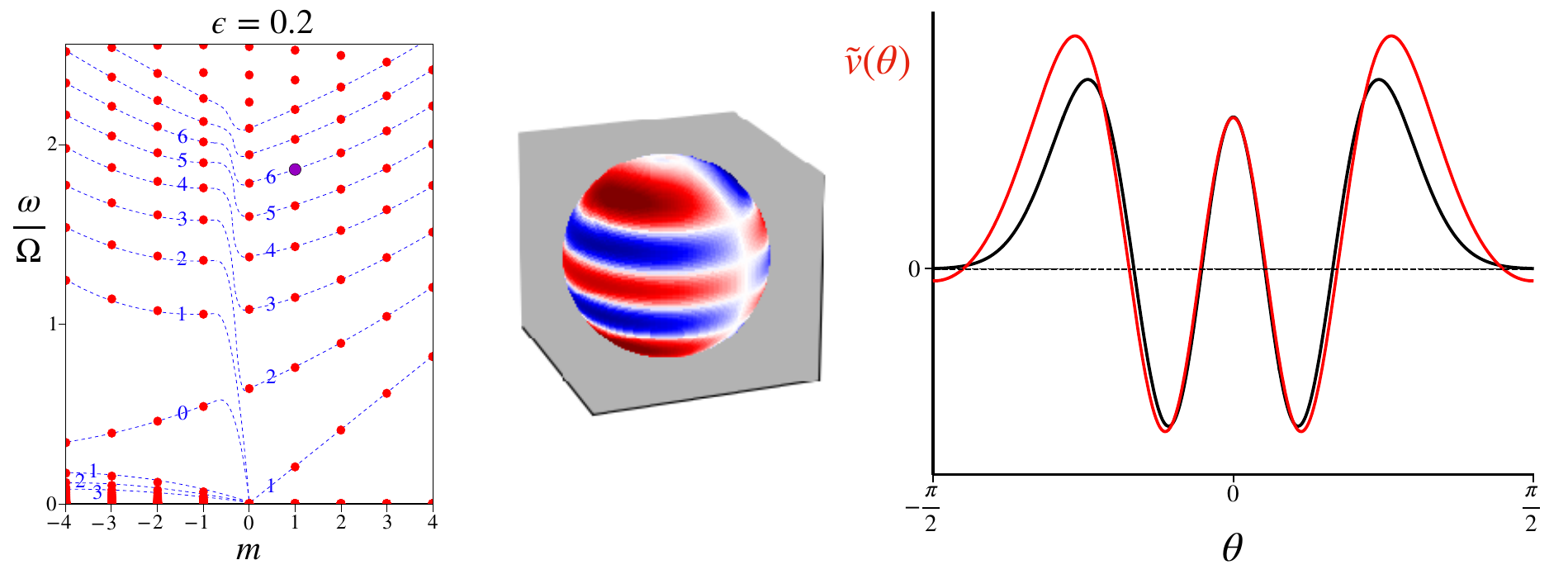}
    \caption{\label{fig:modes_zeros} Additional zeros of $\Tilde{v}$ for $\epsilon = 0.2$. Left: we consider a mode with azimuthal wave number $m=1$ (purple dot). Middle: Meridional velocity $\mathrm{Re} \left( \Tilde{v}(\theta) \ee^{\ii m \phi} \right)$ of the mode. Right: Comparison between the wave function $\Tilde{v}(\theta)$ (red curve) and Expression \eqref{eq:Matsuno_Hermite} for the closest Matsuno mode (black curve). The two wave functions fit well in the Yoshida waveguide but not at higher latitudes, and we notice that the mode on the sphere has $p=6$ zeros whereas the Matsuno mode has $p=4$ zeros.}
\end{figure}

All numerical solutions of the shallow-water eigenvalue problem throughout the paper are calculated with the spectral solver \texttt{Dedalus}. Our scripts can be found at \url{https://github.com/ArmandLeclerc/STArWaRS}. The red dots in the spectra of Figures \ref{fig:asymptotic_regimes}, \ref{fig:Rossby_spherical}, \ref{fig:transition}, \ref{fig:branches_v_label}, \ref{fig:modes_0.2}, \ref{fig:modes_2} and \ref{fig:modes_zeros} are obtained by projecting Equations \eqref{eq:Fourier_SW_bis} on the basis of spin-weighted harmonics implemented in the $3^{\rm rd}$ version of \texttt{Dedalus} \cite{vasil2019tensor}. In the form \eqref{eq:Fourier_SW_bis}, the eigenvalue problem is solved on the sphere, using $128$ harmonics for each value of the azimuthal wave number $m$. In contrast, the blue dots in the plots of Figure \ref{fig:Rossby_spherical} are obtained by expressing the wave equations with the meridional coordinate $\theta$, and projecting the wave functions on a basis of $128$ Chebyshev polynomials of the variable $\theta \in (-\pi / 2,\pi / 2)$, which was already implemented in the earlier version of the solver \cite{burns2020}. The Rossby frequencies obtained with both methods are superimposed in Figure \ref{fig:Rossby_spherical}, for comparison. While solving the wave equations on the sphere only requires imposing the regularity of the velocity and height perturbation fields (this automatically implies that $m$ takes integer values), solving Equation \eqref{eq:Schrodinger_SW} can be done formally for any value of $m$ and requires imposing one Dirichlet boundary condition at each pole, i.e. $\theta = \pm \pi / 2$. We chose the conditions $\ii m \Tilde{u} \mp \Tilde{v} = 0$ at $\theta = \pm \pi / 2$ to reflect single-valuedness of the velocity at the poles, i.e. $\partial_\phi \Tilde{\mathbf{v}} |_{\theta = \pm \pi / 2} = 0$. 
Note that, with both methods, the value of $\epsilon$ and the number of spectral points are the only parameters that need to be defined.

For small enough $\epsilon$, the frequencies of shallow-water waves on the rotating sphere coincide with those of the Matsuno spectrum. However, the wave function $\Tilde{v}(\theta)$ has two more zeros compared to the Hermite wave functions computed by \cite{matsuno1966}. Owing to the Gaussian trapping of waves at the equator, these additional zeros are barely visible for small values of $\epsilon$, such as in Figure \ref{fig:modes_0.2}, whereas they are appreciable for moderate values of $\epsilon$ (Figure \ref{fig:modes_2}). Figure \ref{fig:modes_zeros} shows the difference with the Matsuno wave function \eqref{eq:Matsuno_Hermite} for a mode with $\epsilon = 0.2$ and $m=1$ (taking $\theta = y/R$ and $\beta = 2\Omega / R$).

\subsection{North-south symmetry of the wave functions} \label{apx:symmetry}

Let us demonstrate that the meridional velocity wave function $v(\theta)$ is either even or odd for any mode. Let us consider the mode of frequency $\omega$, represented by a complex-valued vector field $X = \begin{pmatrix} u(\theta) & v(\theta) & \eta(\theta) \end{pmatrix}^\top$ such that $\mathcal{H} X = \omega X$, where the operator $\mathcal{H}$ is expressed in Equation \eqref{eq:Schrodinger_SW}. Since $f(\theta)$ and $\beta_{\rm g} (\theta)$ are odd functions of the latitude, $\mathcal{H}$ commutes with the composition of parity ($\theta \rightarrow -\theta$) and complex conjugation, which can be written as $\mathcal{P} \, \mathcal{H}(m) = \overline{\mathcal{H}(m)} \, \mathcal{P}$. Applying this equality to the vector $X$, one sees that $\overline{\mathcal{P} X}$ is also an eigenvector of $\mathcal{H}$ for the same frequency $\omega$, which means that there is a scalar $z$ such that $\mathcal{P} X = z \overline{X}$. Additionally, since both parity and conjugation are mathematical involutions, the modulus of $z$ is $1$. Besides, the wave function $v(\theta)$ obeys a second-order differential equation with real coefficients (Equation (3.3) of \cite{muller1995shallow}), therefore $v$ is a real-valued function, up to a constant phase that we choose to be zero. The scalar $z$ is thus equal to $\pm 1$ and $v$ has thus even ($z=+1$) or odd ($z=-1$) parity with latitude.\\

Furthermore, using only the first and third equations of the system \eqref{eq:Schrodinger_SW}, one can express the fields $u$ and $\eta$ in terms of $v$ and its derivative $\dd v / \dd \theta$, and see that these two fields are purely imaginary. This means that the zonal velocity $u$ and height perturbation $\eta$ are in phase quadrature with the meridional velocity $v$, and that they both have even (resp. odd) north-south symmetry if $v$ has odd (resp. even) north-south symmetry (Note that an odd function $v(\theta)$ actually reflects equatorially-symmetric meridional velocity, and conversely even $v(\theta)$ corresponds to anti-symmetric meridional velocity). This symmetry of the wave functions, which is also true for the modes of the unbounded $\beta$-plane, reflects the north-south symmetry of the geometry itself, which is present whether one is dealing with the unbounded $\beta$-plane or the sphere. This symmetry of the wave function $v(\theta)$ also implies that the zeros revealed on the sphere always come in pair.

\subsection{Weyl correspondence and symbol} \label{apx:WW}

Throughout the paper, we naturally use notions such as \textit{scalar product, Hermitian operator} and \textit{Weyl symbol}, assuming an implicit algebraic structure to the fields of the problem. Let us formally define these notions here. First of all, Equation \eqref{eq:Schrodinger_SW}, $\mathcal{H} X = \omega X$ with $X = \begin{pmatrix} u & v & \eta \end{pmatrix}^\top$, implies that we work with a Hilbert space of 3-component vectors, whose components are complex-valued functions of $\theta \in (-\pi / 2 , \pi / 2)$. The matrix operator $\mathcal{H}$ acts on this space. Since the component fields are defined by multiplying the real perturbed velocity components and elevation (up to factors $\sqrt{h_0},\sqrt{g}$) by a metric factor $\sqrt{\cos (\theta)}$, the metric term $\cos (\theta)$ in quadratic integrals over the sphere is absorbed. For instance, the mechanical energy of the perturbation is proportional to
\begin{equation}
    \int_{-\frac{\pi}{2}}^\frac{\pi}{2} \dd \theta \left( |u|^2 + |v|^2 + |\eta|^2 \right) \ ,
\end{equation}
which is a conserved quantity of the shallow-water equations. This leads to the following definition of the scalar product:
\begin{equation}
    \langle X_1 , X_2 \rangle = \int_{-\frac{\pi}{2}}^\frac{\pi}{2} \dd \theta \: X_1^\dagger X_2 \ ,
\end{equation}
where $^\dagger$ stands for the transposition-conjugation of vectors and matrices. The operator $\mathcal{H}$ is thus Hermitian for this scalar product, i.e. for any $X_1, X_2$, the equality
\begin{equation}
    \langle X_1 , \mathcal{H} X_2 \rangle = \langle \mathcal{H} X_1 , X_2 \rangle \ ,
\end{equation}
is satisfied. In fine, absorbing the metric term $\sqrt{\cos (\theta)}$ in the definition of the fields re-defines the scalar product in curved space as a Cartesian scalar product. This allows us to use the definition of the \textit{Weyl correspondence} (or \textit{Wigner map}) in flat spatial coordinates (i.e. without necessity to introduce an elaborated Stratonovich-Weyl operator kernel \cite{gneiting2013quantum}), which connects the wave operator $\mathcal{H}$ of Equation \eqref{eq:Schrodinger_SW} to its Weyl symbol $H$, defined by Expression \eqref{eq:symbol}, through the bijective map
\begin{equation}
    \mathcal{H} X(\theta) = \frac{1}{2 \pi} \int_{-\frac{\pi}{2}}^\frac{\pi}{2} \dd \theta' \int_{-\infty}^{+\infty} \dd p_\theta \: \ee^{\ii p_\theta (\theta - \theta')} \: H \left( p_\theta , \frac{\theta + \theta'}{2} \right) X(\theta') \ ,
\end{equation}
where we have used the conjugated momentum of $\theta$, $p_\theta = (R/c) K_\theta$. $H$ is a function of $\theta$ and $p_\theta$, however we rather use the variable $K_\theta$ in the paper, to avoid the $c/R$ term. $H$ is the matrix of Weyl symbols of the components of the matrix operator $\mathcal{H}$. Those are either products with functions $F(\theta)$ (including $M$) or the derivative $-\ii \dd / \dd \theta$, whose symbols are
\begin{equation} \label{eq:Wigner_scalar}
    F(\theta) \cdot \longrightarrow F(\theta) \quad \text{and} \quad -\ii \frac{\dd \cdot}{\dd \theta} \longrightarrow p_\theta \ .
\end{equation}

An important property of the Wigner map is that an operator is self-adjoint if, and only if its Weyl symbol is a Hermitian matrix \cite{hall2013quantum}, or just a real-valued function for scalar operators such as those of relations \eqref{eq:Wigner_scalar}.

\subsection{Definition and computation of the Chern numbers} \label{apx:topology}

\begin{figure}[H]
    \centering
    \includegraphics[scale=0.5]{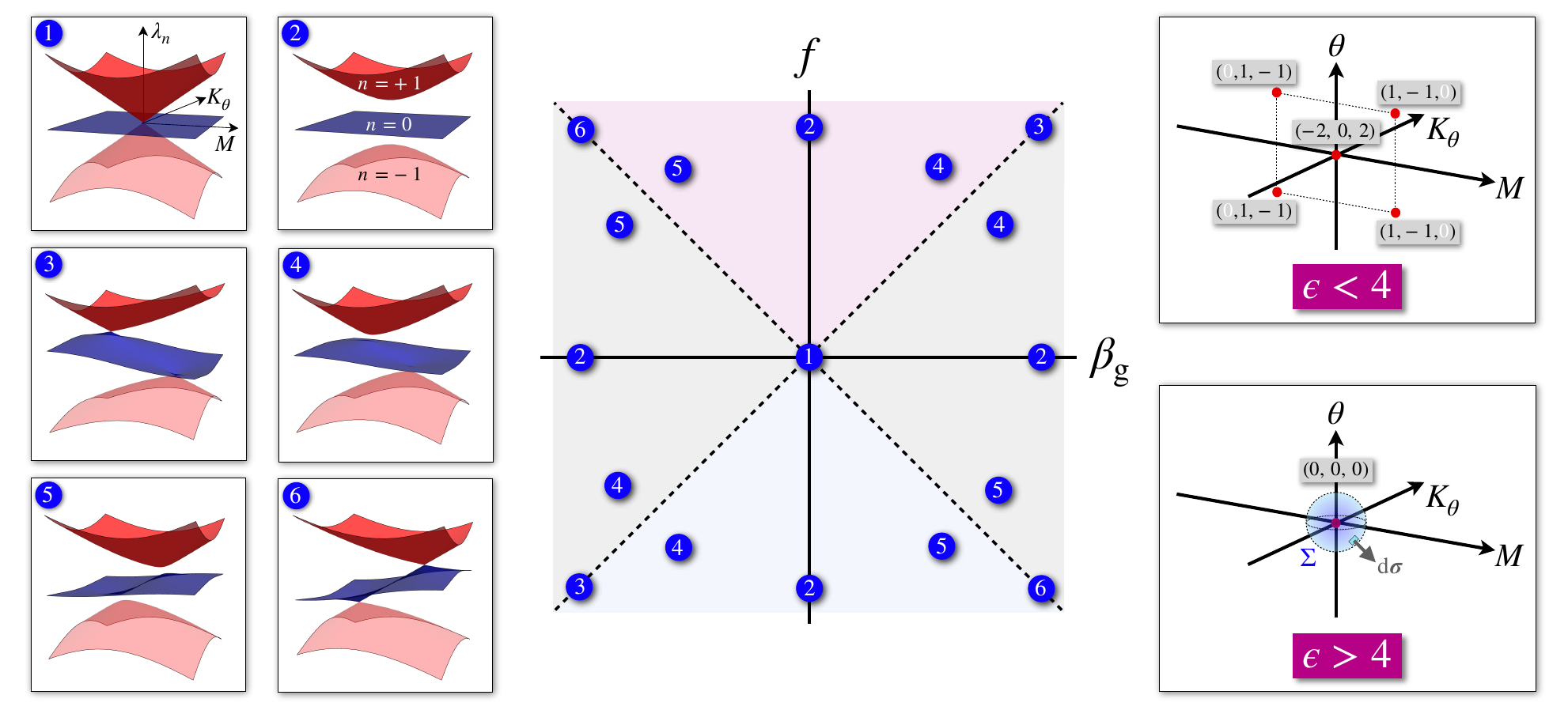}
    \caption{\label{fig:DP_diagram} Eigenbands and Chern numbers of the symbol \eqref{eq:symbol}. Left: eigenvalues of $H$ as functions of $(M,K_\theta)$, for different values of $f$ and $\beta_{\rm g}$ (which can be formally considered as independent parameters, although they really are both functions of the latitude $\theta$). The middle panel provides the corresponding locations of $(f,\beta_{\rm g})$. The dashed lines $f=\pm \beta_{\rm g}$ indicate where the symbol has multiple eigenvalues. Spectra 1 and 3 are the local dispersion relations achieved for the actual prescribed functions $f(\theta),\beta_{\rm g}(\theta)$. Right: Position and Chern numbers $(C_{-1},C_0,C_{+1})$ of the degeneracy points of the symbol in parameter space $(M,K_\theta , \theta)$, for $\epsilon<4$ and $\epsilon>4$. The white zeros indicate that the corresponding eigenband is not involved in the degeneracy (integrating the Berry curvature on a closed surface in which there is no degeneracy point yields zero). The blue sphere is an example for the integration surface $\Sigma$ of definition \eqref{eq:Chern_integration}.}
\end{figure}

\begin{figure}[H]
    \centering
    \includegraphics[scale=0.45]{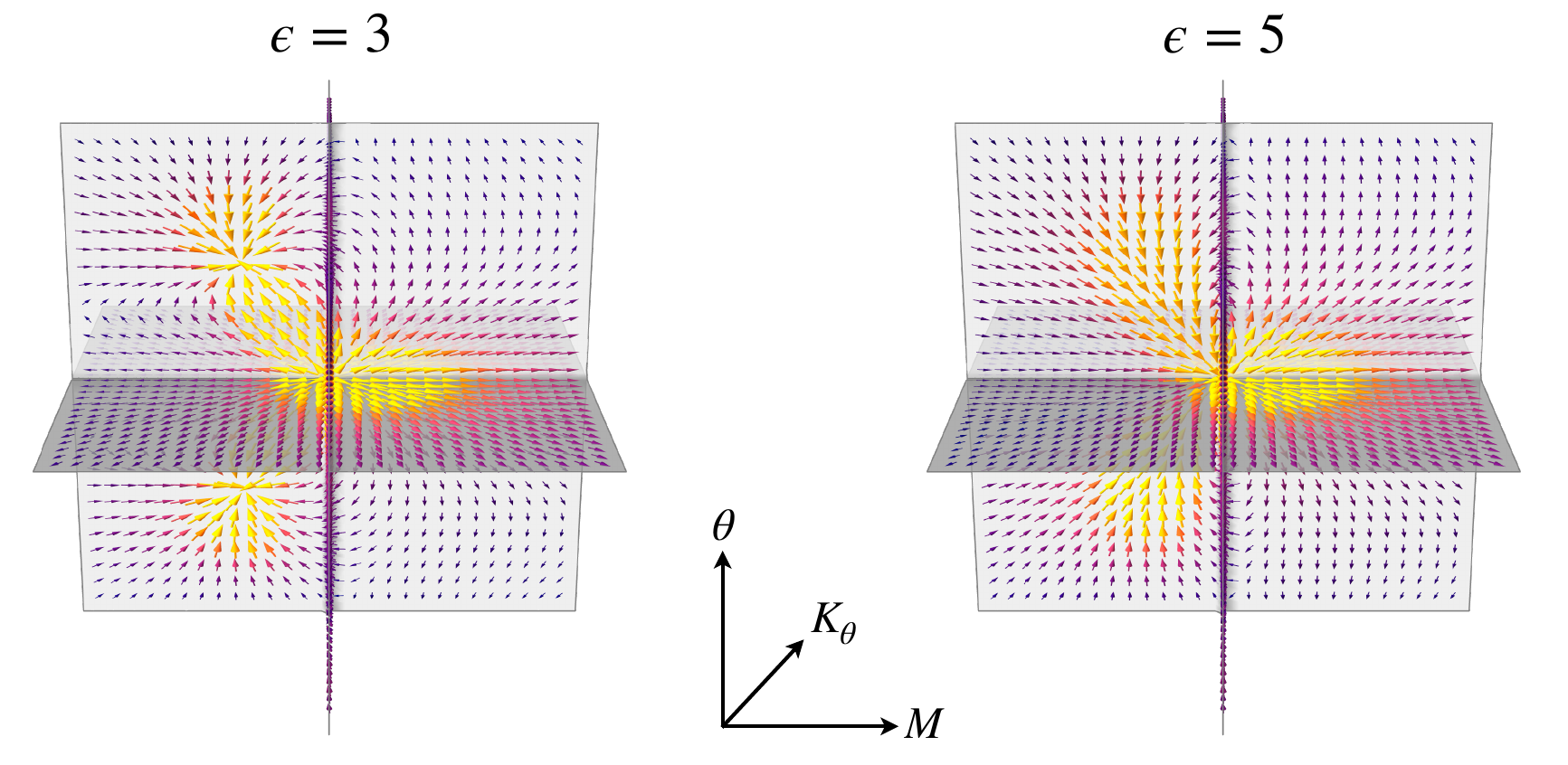}
    \caption{\label{fig:Berry_curvature} Berry curvature $\mathbf{F}_n$ of the eigenband $n=+1$ of the Weyl symbol $H$. For $\epsilon < 4$ (left), there are three topological charges, which merge into a single non-charged degeneracy point for $\epsilon > 4$ (right). The center of each plot is the point $(M,K_\theta,\theta)=(0,0,0)$.}
\end{figure}

The Chern numbers of the degeneracy points of $H$ are computed in the same way as in \cite{delplace2017,perrot2019topological,venaille2021,perez2022unidirectional,leclerc2022topological,perez2022topological}. For a degeneracy point involving $\mu$ eigenbands (i.e. corresponding to an eigenvalue of $H$ of multiplicity $\mu$), $\mu$ Chern numbers $C_n$ are associated ($n$ is the band index), which can be computed as the flux of the \textit{Berry curvature} of band $n$ through a closed surface enveloping the degeneracy point in the 3-dimensional parameter space $(M , K_\theta, \theta)$ (see Figure \ref{fig:DP_diagram}):
\begin{equation} \label{eq:Chern_integration}
    C_n = \frac{1}{2 \pi} \iint_\Sigma \dd \boldsymbol{\sigma} \cdot \mathbf{F}_n \ ,
\end{equation}
where $\Sigma$ is an integration surface enclosing the degeneracy point in parameter space, and $\dd \boldsymbol{\sigma}$ an element of surface with the outward orientation. Note that using either $m$ or $M$ (which is a function of both $m$ and $\theta$, strictly speaking) does not change the value of the Chern numbers. We chose the variable $M$ for convenience. Besides, the definition \eqref{eq:Chern_integration} implies that $m$ is formally considered as a continuous variable of the symbol $H$, which is not a problem as far as the definition \eqref{eq:symbol} is concerned. However, when it comes to finding the modes of $\mathcal{H}$, $m$ must be an integer so that those are regular and single-valued on the sphere. The Berry curvature \cite{berry1984quantal} is defined as
\begin{equation} \label{eq:Berry}
    \mathbf{F}_n = \boldsymbol{\nabla} \times \left( \ii \Psi_n^\dagger \boldsymbol{\nabla} \Psi_n \right) = -\mathrm{Im} \left( \sum_{n' \neq n} \frac{\Psi_n^\dagger (\boldsymbol{\nabla} H) \Psi_{n'} \times \Psi_{n'}^\dagger (\boldsymbol{\nabla} H) \Psi_{n}}{(\lambda_n - \lambda_{n'})^2} \right) \ ,
\end{equation}
where the complex vectors $\Psi_n$ are the normalised eigenvectors of $H$ for the eigenvalue $\lambda_n$, and the operator $\boldsymbol{\nabla}$ (resp. $\boldsymbol{\nabla} \times$) is the gradient (resp. curl) on the parameter space $(M , K_\theta, \theta)$. Expressions \eqref{eq:Berry} reveal essential properties of the Berry curvature: it is a 3-component, real-valued vector field defined over the parameter space, for each band. It is smaller at points where the eigenvalue $\lambda_n$ is separated from the others, and it diverges at degeneracy points of the eigenband $n$. For any other point, the sum of the Berry curvatures of all eigenbands is zero, thus, for a given degeneracy point, the sum of the Chern numbers is zero as well. By analogy with the electric field produced by a static charged particle, the topological charge $C_n$ of a degeneracy point generates a Berry curvature $\mathbf{F}_n$ which diverges for positive Chern number, and converges for negative Chern number (see Figure \ref{fig:Berry_curvature}). The Berry curvature $\mathbf{F}_n$ is otherwise divergence-free, i.e. its flux through any closed surface that does not contain a degeneracy point of the eigenband $n$ is zero. In terms of topology, the Chern number $C_n$ of a degeneracy point $P$ characterises and quantises the existence of singularities of the phase (i.e. topological defects) of the complex vector $\Psi_n$ when one attempts to continuously define it around $P$ (see e.g. \cite{perez2022topological}, section 2.3). 
The Berry curvature of the inertia-gravity eigenband $n=+1$ is represented in Figure \ref{fig:Berry_curvature}, for $\epsilon < 4$ and $\epsilon > 4$.

\begin{figure}[H]
    \centering
    \includegraphics[scale=0.45]{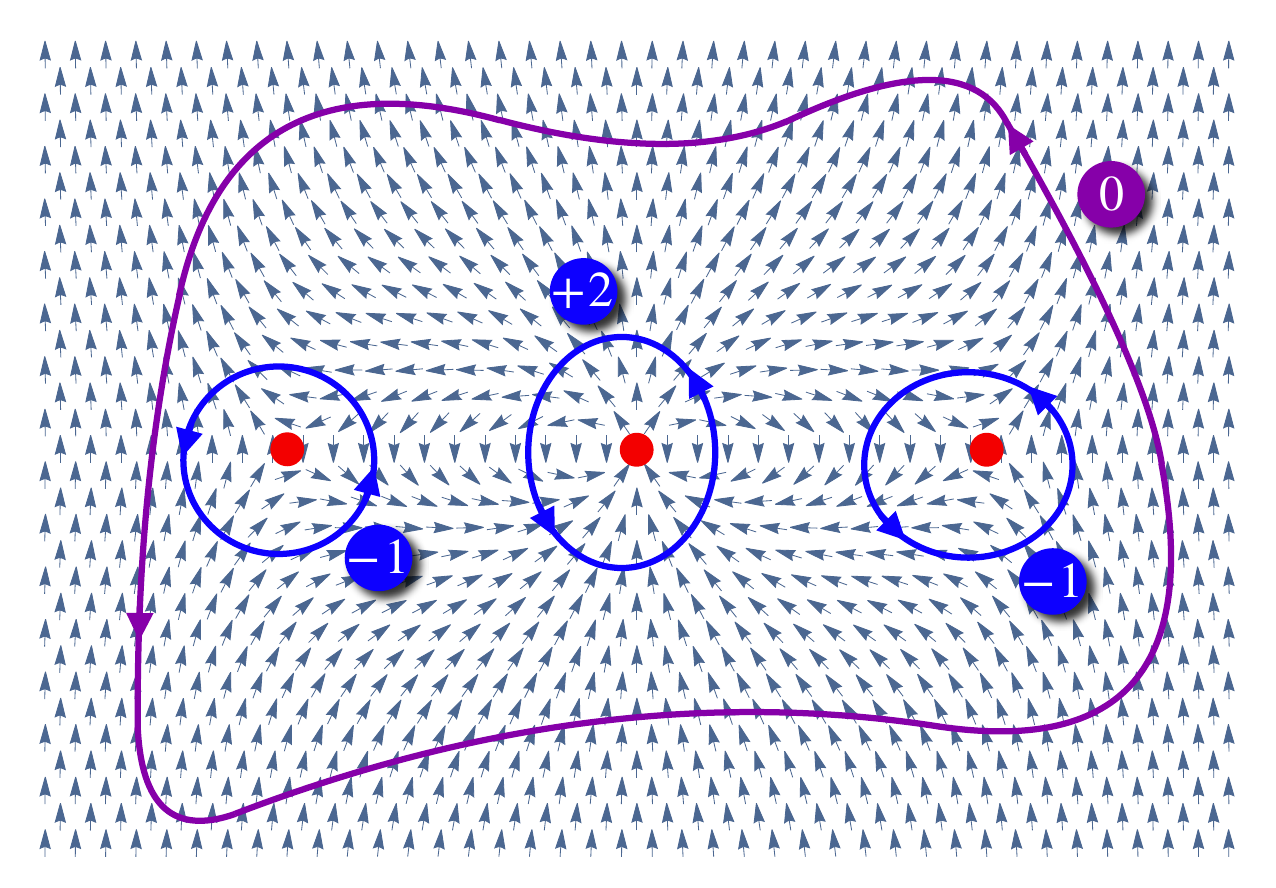}
    \caption{\label{fig:topological_defects} Topological defects of a tangent vector field over the torus (square with periodic boundary conditions), whose winding numbers compensate, in virtue of the Gauss-Bonnet theorem (the Euler characteristic of the torus is 0). The integers are formally defined as the integral of $\dd \xi / 2 \pi$ along the corresponding path, where $\xi$ is the algebraic angle between the horizontal direction and the vectors. A topological defect having non-zero winding number is characteristic of the fact that such a function $\xi$ cannot be continuously defined around a defect (blue curves). However, it can be along a path that winds all three defects of compensating charges (purple curve). This graphic situation illustrates the topology of the shallow-water spectrum on the sphere in the case $\epsilon < 4$: there are several degeneracy points with non-zero Chern numbers which add up to zero for each eigenband. The vector field can be continuously deformed until the three charged topological defects merge into a topologically neutral point.}
\end{figure}

The implication of the Chern numbers $C_n$ on the topology of the complex-valued vectors $\Psi_n$, defined over the 3-dimensional parameter space, is analogous to the quantisation of the topological defects of a smooth, real-valued vector field tangent to a surface (see e.g. \cite{perez2022topological}, section 2.3). For each point where the vector field vanishes, one can define an integer which quantises the winding of the vector field around 0, following a closed path around this point in a counter-clockwise direction (see Figure \ref{fig:topological_defects}). In this case, the sum of such numbers is equal to the Euler characteristic of the surface, in virtue of the Gauss-Bonnet theorem. However, this analogy cannot be considered as an equivalence, since we are dealing with complex-valued vector field in a 3-dimensional parameter space, instead of real-valued vector field defined over 2-dimensional manifolds.

As far as computation of the Chern numbers is concerned, the 3-by-3 matrix $H$ is simple enough so it can be done by direct numerical integration, which was performed with \texttt{Mathematica} using Expressions \eqref{eq:Berry} and \eqref{eq:Chern_integration}. Our script can be found at \url{https://github.com/ArmandLeclerc/STArWaRS}. The values of the Chern numbers, given in the paper, are summarised in Figure \ref{fig:DP_diagram}. For symbol matrices of higher dimension, such direct computation quickly becomes difficult, however other numerical methods are possible, adapted for instance from \cite{fukui2005chern}.

\end{document}